\def \half {{1\over 2}}
\newcommand{\sss}{{\scriptscriptstyle }}

\def\roughly#1{\mathrel{\raise.3ex\hbox{$#1$\kern-.75em
\lower1ex\hbox{$\sim$}}}}
\def\lsim{\roughly<}
\def\gsim{\roughly>}
\def\pref#1{(\ref{#1})}
\def\nn{\nonumber}
\def\eq{\begin{equation}}
\def\eeq{\end{equation}}
\def\eqa{\begin{eqnarray}}
\def\eeqa{\end{eqnarray}}
\def\cG{{\cal G}}

\def\bp{{\bf p}}
\def\scA{{\cal A}}
\def\scB{{\cal B}}
\def\scC{{\cal C}}
\def\scD{{\cal D}}
\def\scE{{\cal E}}
\def\eps{\epsilon}
\def\veps{\varepsilon}
\def\dsl{\hbox{/\kern-.6600em$D$}}
\def\psl{\hbox{/\kern-.6600em$P$}}
\def\qsl{\hbox{/\kern-.5600em$Q$}}
\def\ksl{\hbox{/\kern-.5600em$k$}}
\def\g{\gamma}
\def\sss{\scriptscriptstyle}
\def\a{\alpha}
\def\phm{{\phantom{-}\,}}
\def\phl{\; {\phantom{\cal L}}}

\documentstyle[prl,aps,epsf]{revtex}
\begin{document}
\twocolumn[\hsize\textwidth\columnwidth\hsize\csname@twocolumnfalse%
\endcsname
\rightline{McGill-01/26, UW/PT 01-27, {\tt hep-th/0201082}} \vspace{3mm}

\draft
\title{Loop-Generated Bounds on Changes to the Graviton Dispersion Relation}

\author{C.P.~Burgess,${}^{a}$ J. Cline,${}^a$, E. Filotas${}^a$,
J. Matias,${}^b$ and G.D. Moore${}^c$}

\address{${}^a$ Physics Department, McGill University,
3600 University Street, Montr\'eal, Qu\'ebec, Canada H3A 2T8.\\
${}^b$ IFAE, Universitat Aut\`onoma de Barcelona, Spain.\\
${}^c$ Department of Physics, University of Washington, Seattle, WA 98195, USA.}
\maketitle

\begin{abstract}
{We identify the effective theory appropriate to the propagation
of massless bulk fields in brane-world scenarios, to show that the
dominant low-energy effect of asymmetric warping in the bulk is to
modify the dispersion relation of the effective 4-dimensional
modes. We show how such changes to the graviton dispersion
relation may be bounded through the effects they imply, through
loops, for the propagation of standard model particles. We compute
these bounds and show that they provide, in some cases, the
strongest constraints on nonstandard gravitational dispersions.
The bounds obtained in this way are the strongest for the fewest
extra dimensions and when the extra-dimensional Planck mass is the
smallest. Although the best bounds come for warped 5-D scenarios,
for which $M_5$ is $O(\hbox{TeV})$, even in 4 dimensions the
graviton loop can lead to a bound on the graviton speed which is
comparable with other constraints.}
\end{abstract}
]

%
%
%%%%%%%%%%%\narrowtext  %<- CLH to get two columns a la PRD
% aim for around 700 lines

\section{Introduction}

Although physics in more than the traditional four dimensions has been
long speculated to be important for describing phenomena at energies
above the electroweak scale, the idea is presently enjoying additional
scrutiny because of two more recent developments.

First came the
understanding of the strong-coupling limit of string theory in terms
of 11-dimensional supergravity interacting on spaces with boundaries
\cite{HW}. Together with the realization that observable particles
can be trapped on these boundaries, or on D-branes, this understanding
freed the string scale from the Planck mass, allowing it to be
as low as the weak scale \cite{ADD} or at intermediate or
grand-unified scales \cite{OtherScales}. Such low string scales
allow extra dimensions to be comparatively large, and so (potentially)
to have much richer implications at experimentally accessible energies.

Second came the observation that even very small extra dimensions
might also have interesting low-energy implications if their geometries
are `warped' \cite{RSI,RSII}. In this framework, the four-dimensional metric
has a nontrivial dependence on position in the extra dimensions,
allowing four-dimensional properties like masses and couplings to
depend in an interesting way on an observer's position within the
extra dimensions.

Although not required by either approach, the broader class of
geometries which are allowed once observable particles are
confined to a brane includes many configurations which violate
Lorentz invariance within the observable four dimensions by
picking out a preferred frame \cite{Lorentz,DRV}, such as by
having gravitating objects displaced from our brane in the extra
dimensions. For example, within the warped framework one can have
a 5D line element of the form
\eq \label{ads}
    ds^2 = -(\a+\Delta) dt^2 + \a d\vec x^{\,2} + {dr^2\over
    \a+\Delta}
\eeq
where $t,$ $\vec{x}$ are 4 dimensional time and space, $r$ is the
extra dimension, $\a(r) = r^2/l^2$ and $\Delta(r) = - \mu/r^2 +
Q^2/r^4$. If $\mu=Q=0$ this is anti-de Sitter space with curvature
radius $l$, and there is no Lorentz violation.  But if $\mu\neq
0$, it is the AdS-Reissner-N\"ordstrom metric with a singularity
at $r=0$ and a speed of light which varies with $r$ as $c^2(r)=1+
{\Delta\over\a}$. There is a preferred frame with 4-velocity
$u^\mu = (1,0,0,0)$ in these coordinates.  In this example, the
preferred frame is not due to visible matter, but rather the
presence of a ``black brane'' at $r=0$ which is displaced from our
brane in the extra dimension. {\it A priori} its effects need not
be small, and could cause observable phenomena.

Very strong limits exist on the size of various Lorentz-violating
effects involving ordinary particles \cite{LVBounds,Glashow}, so
one expects these to provide the most stringent tests of any
Lorentz-violation effects predicted in the brane-world scenario.
Although this expectation is borne out, the surprise in these
scenarios is always the comparative difficulty of finding good
constraints in the event that the only known particle living in
the bulk is the graviton, since so little is directly known about
the graviton's properties. Some comparatively weak limits do
exist, coming from Parameterized Post-Newtonian tests of General
Relativity within the solar system \cite{WillBook,Will}, and from
its success in describing the energy loss of the binary pulsar,
1913+16 \cite{Will,Taylor,Carlip}.

Much stronger limits on changes to the graviton dispersion
relation may also be obtained if they permit high-energy particles
to \v Cerenkov radiate gravitons. In this case bounds may be obtained
from our observation of very high energy particles in cosmic rays.
Assuming these cosmic rays to be protons -- as the best evidence
indicates \cite{protvsphot} -- the constraint $(c_p - c_g)/c_p <
10^{-15}$ was obtained in this way \cite{MN}, strongly restricting the
case where the graviton velocity, $c_g$, is smaller than that of the
proton, $c_p$.

Our purpose in this paper is to provide a complementary constraint
for the case where protons cannot radiate gravitons, and so the
above bound does not apply. We obtain our constraint by computing
some of the Lorentz-violating effects which are implied for
fermions and photons by radiative corrections involving gravitons.
Because the bounds on such corrections for electrons and photons
are extremely good, we are able to infer reasonably strong bounds
on Lorentz violation in the graviton sector.

We find two kinds of effects, each of which depends on physics at
different scales. We find the one-loop graviton-induced
contributions to particle phase velocities to be highly suppressed
by the mass of the particle involved. This sensitivity is stronger
the higher the number of dimensions the graviton sees. The photon
and electron are particularly insensitive to this kind of
correction. Because they are so mass-dependent, the largest
contributions to light particles are often due to higher-loop
graphs within which a one-loop graviton-induced modification to
the propagation of a heavy particle (like a $W$ boson or top
quark) is inserted. This makes predictions sensitive to the
ultraviolet structure of the theory.

Graviton loops also change light-particle dispersion relations
by introducing contributions which involve higher powers of the
particle momentum. We compute these and find that they
can be generically less ultraviolet sensitive
than are changes to particle phase velocities. Interestingly, the
low-energy contributions of this form ({\it i.e.} the contributions
due to the massless 4D graviton) are already large enough to provide
interesting bounds.

We organize our presentation as follows. In the next section we
discuss in general terms how Lorentz-violating effects enter into
the low-energy 4D theory at energies well below the
compactification scale. This is the regime of interest for
experiments, and is {\it not} the regime in which one can consider
bulk particles to be moving on ballistic trajectories through the
extra dimensions. Then in section III we discuss the fermion and
photon self energies in general, to identify which features are
required in order to make comparison with experiments. Section IV
follows with the calculation of the graviton loop, and the
derivation of the induced Lorentz violation amongst visible
brane-based particles. Given the mass dependence of the results
of Section IV, Section V gives conservative estimates as to the
size which might be expected for photons, electrons and protons.
Section VI describes the bounds on Lorentz violation for these
particles, and computes the bound which may be inferred indirectly
on the strength of Lorentz violation in the gravity sector. Our
conclusions are summarized in section VI, and some of the
cumbersome intermediate results are gathered into two appendices.

\section{The 4d Effective Picture}

In the brane-world picture photons and electrons usually are
constrained to move on our four-dimensional brane, while gravitons are
free to explore the higher-dimensional bulk surrounding space. It is
intuitive in this kind of picture that brane and bulk particles might
propagate differently, since bulk particles might be free to take
`shortcuts' through the extra dimensions which are forbidden to
brane-bound particles \cite{Shortcuts}. This possibility has been
proposed to be, in some instances, a virtue in that it may provide a
novel way to address some cosmological problems \cite{DRV,CosmoShort}.

Although this kind of ballistic picture is appealingly intuitive,
it is not really appropriate for the low energies at which tests of
the dispersion relation actually take place. Experimental tests only involve
the one graviton which we see in 4 dimensions, and so only involve
the very lowest Kaluza-Klein (KK) state. By contrast, assigning the graviton
localized trajectories in the extra dimensions presupposes a localized
wave packet in these dimensions, which cannot be constructed purely
from the lowest-energy mode.

We briefly detour in this section to describe how extra-dimensional
Lorentz-violation appears in the low-energy effective Lagrangian
which describes the lowest KK mode. Although we start by using the
simplest example of a scalar field, the conclusions we draw will be
shown to be equally valid for higher spin fields.

\subsection{A Scalar Field}
Consider, therefore, a bulk scalar field, $\Phi(x,y)$, where $x^\mu$
labels the 4 dimensions parallel to the brane, and $y^m$ labels
the $n$ various transverse dimensions. The four-dimensional field
content is obtained by resolving $\Phi$ into a basis of modes
in the extra dimensions:
\eq
\Phi(x,y) = \sum_k \varphi_k(x) \, u_k(y),
\eeq
where the basis functions, $u_k(y)$, are eigenfunctions of the
appropriate kinetic/mass operator: $\Delta u_k = \omega_k \, u_k$.

The kinetic 4D action for the KK modes, $\varphi_k(x)$, is
obtained by inserting the mode expansion into the higher-dimensional
action and integrating the result over the
extra dimensions. Using an assumed form for the extra-dimensional
background metric:
\eq
\cG_{MN} = \pmatrix{ g_{\mu\nu}(x,y) & 0 \cr
0 & h_{mn}(y) \cr}
\eeq
one finds in this way:
\eqa
S &=& -\, \frac12 \int d^{\,4}x \, d^{\,n}y \; \sqrt{-\cG} {\cG}^{MN}
\partial_M \Phi \partial_N \Phi \nonumber\\
&=& - \, \frac12 \int d^{\,4}x \; \sqrt{-G} \,
G^{\mu\nu}_{kl} \; \partial_\mu \varphi_k \partial_\nu \varphi_l
+ \cdots
\eeqa
where $\cG = \det{\cG_{MN}}$, {\it etc.,} and the effective
four dimensional metric governing the kinetic terms is
\eq
\label{EffG}
\sqrt{-G}\, G^{\mu\nu}_{kl}(x) := \int d^{\,n} y\;
\sqrt{-gh}\,  g^{\mu\nu}(x,y) \; u^*_k(y) \,
u_l(y).
\eeq

The main point is that the metrics (plural since there is a
different metric for each choice of the indices $k,l$) defined by
eq.\ \pref{EffG} differ, in general, from the induced metric on
the brane, $\gamma_{\mu\nu}$, which appears in the kinetic term
for fields which are trapped on a brane. For instance, for a brane
defined by the surface $y=y_0$ the induced metric is simply
$\gamma_{\mu\nu} = g_{\mu\nu}(x,y_0)$. This is ultimately the
source of Lorentz-violating effects due to the bulk metric.

If we focus purely on the massless four-dimensional
mode, which we label by $k=0$, then we must integrate
out the other, more massive, KK modes. The kinetic term 
for the massless field,
$\varphi = \varphi_0(x)$, contains the metric
$G^{\mu\nu} = G^{\mu\nu}_{00}$ which may differ
from the induced metric on the brane, $\gamma^{\mu\nu}$.
In the rest of this paper we will focus on the implications
of this difference.

Two limits of the metric $G^{\mu\nu}$ bear highlighting.
First, in the absence of warping of the bulk metric
({\it i.e.,} if $g^{\mu\nu}(x)$ is independent of $y$)
eq.\ \pref{EffG} becomes:
\eqa
\sqrt{-G}\, G^{\mu\nu}_{kl} &=& \sqrt{-g} \, g^{\mu\nu}(x)
\int d^{\,n}y \; \sqrt{h} \, u^*_k \, u_l \nn\\
&=& \delta_{kl} \; \sqrt{-g} \, g^{\mu\nu}(x), \eeqa
where the second line uses the orthonormality condition
of the basis modes, $u_k(y)$. In this case bulk and
brane modes see the same metric, for branes defined by
surfaces $y=y_0$.

Second, in the absence of a preferred frame in the
bulk metric (such as the AdS metric used by Randall
and Sundrum) the metrics $G_{\mu\nu}$ and $\gamma_{\mu\nu}$
must be conformal to one another ({\it i.e.,}
$G_{\mu\nu}(x) = f(x) \, \gamma_{\mu\nu}(x)$),
since Lorentz invariance implies they must both locally
be proportional to the Minkowski space metric $\eta_{\mu\nu}$.

\subsection{The 4D Graviton} A similar story gives the influence of
Lorentz-violating effects on the propagation of low-energy gravitons
within the effective theory below the compactification scale (or the
AdS curvature scale, in the case of RS-II-like models without
compactification \cite{RSII}), although the details are a bit more
complicated due to gauge invariance. Our purpose here is to identify
the leading contributions of Lorentz violations in the higher-energy
theory to graviton propagation in the low-energy 4D theory, and to show
that they always may be cast in terms of an appropriate shift of the
background metric.\footnote{We thank M. Pospelov for helpful
conversations on this point.}

We assume: ($i$) our interest is in energies very low compared to
the compactification scale, allowing a treatment in terms of the
low-energy 4D effective theory; ($ii$) there is only a single
massless spin-2 graviton mode in this effective theory, and
($iii$) that the dominant effect of the higher-dimensional theory
is to break Lorentz invariance but not translation invariance or
rotational invariance (in the preferred frame). Under these
circumstances the 4D effective theory
involves an effective 4D metric field coupled to an order
parameter, $u_\mu$, which defines the preferred frame. The
assumptions of unbroken rotational invariance imply $u_\mu$ is
timelike, and we rescale it so that it is normalized, satisfying
$g^{\mu\nu} u_\mu u_\nu = -1$.

On grounds of general covariance, the 4D effective theory with
the fewest derivatives has the form:\footnote{Our metric is
`mostly plus', and we follow Weinberg's curvature conventions
\cite{GravCosm}.  Momentum 4-vectors are upper case ($P$),
the spatial vector is boldface (${\bf p}$), the magnitude of the spatial
piece is lowercase ($p$).}
\eq \label{d4DEH} {\cal L} = -\, {1 \over 2 \kappa_4^2} \;
\sqrt{-g} \Bigl[ R + a \; R^{\mu\nu} \, u_\mu u_\nu \Bigr] +
\dots \, ,\eeq
for some dimensionless constant $a$. Here the ellipses indicate
higher-derivative terms, and we have not written a cosmological
constant term, which we assume to be negligible. (We shed no light
in this paper on the vexing cosmological constant problem.) 
Here $\kappa_4^2 = 8 \pi \, G_4 = 1/M_4^2$, where
$G_4$ and $M_4$ are the usual 4D Newton's constant and
(rationalized) Planck mass: $M_4 \sim 2 \times 10^{18}$ GeV. In
order of magnitude we expect $a \sim \kappa^2_4 \Lambda^2$, where
$\Lambda$ is the scale associated with the Lorentz-violating
effects at higher energies. Clearly $a$ is naturally very small to
the extent that $\Lambda$ is much smaller than the 4D Planck
scale.

The main point now follows. The second term of eq.\ \pref{d4DEH}
may be completely absorbed by performing the following field
redefinition:
\eq \label{fieldredef} g_{\mu\nu} \to g_{\mu\nu} - \, {a \over 2}
\; \Bigl[ g_{\mu\nu} + 2 \, u_\mu u_\nu \Bigr] . \eeq
After having performed this redefinition (and a constant rescaling
of the metric), graviton fluctuations about flat space are
described performing the expansion $g_{\mu\nu}(x) = G_{\mu\nu} + 2
\kappa_4 \, h_{\mu\nu}(x)$, in eq.\ \pref{d4DEH}, using
\eq \label{metricform} G_{\mu\nu} = \eta_{\mu\nu} - \delta c_g^2
\, u_\mu u_\nu, \eeq
where $\eta_{\mu\nu} = \hbox{diag} (-1,1,1,1)$ is the usual
Minkowski metric, $u_\mu = \eta_{\mu\nu} u^\nu$, and $\delta
c_g^2$ is a small quantity which we shall see has the
interpretation as a change in the maximum propagation speed,
$c_g$, of the graviton. The field $h_{\mu\nu}$ here is the
canonically-normalized field describing graviton propagation.

Just as was the case for the scalar field, the leading effect of
higher-energy Lorentz violation is in this way seen to be a
modification of the background metric through which the graviton
propagates.

As a concrete illustration of this general argument, we can
compute the effective 4-D gravitational metric corresponding to
eq.\ (\ref{ads}), treating the Lorentz-violating term as a
perturbation.  The 4D effective gravitational action can be
obtained by expanding the 5D action $S$ to linear order in $\delta
g_{\mu\nu} = (-\mu/r^2+Q^2/r^4)u_\mu u_\nu$, and integrating over
the extra dimension.  If we write
\eq
    ds^2 \cong {r^2\over l^2}
    \left( g_{\mu\nu}(x) - {l^2\over r^2}\delta g_{\mu\nu}
    \right) dx^\mu dx^\nu +  {l^2\over r^2} dr^2
\eeq
(the correction $\Delta(r)$ in $g_{rr}$ can be neglected to leading order
in $\Delta$) then\footnote{For a warped metric of the form $ds^2 =
a(r)g_{\mu\nu}dx^\mu dx^\nu + b(r) dr^2$, the reduction of the
gravitational action from 5D to 4D is $S = -\frac12 M_5^3
(\int a\sqrt{b}\,dr)\sqrt{-g}R$, where $R$ is the Ricci scalar constructed
from the 4D part of the metric, $g_{\mu\nu}$.}
\eqa
    \delta  S &=& \frac12 M_5^3 \left(\int_{r_1}^{r_2} dr\,
    {r\over l}\cdot{l^2 \over r^2}\left[-{\mu\over r^2}+{Q^2\over r^4}
    \right]\right)
    \nonumber\\
    &&\times\sqrt{-g}\left(R^{\mu\nu}-\frac12 g^{\mu\nu}R\right) u_\mu u_\nu
\eeqa
The upper limit of integration corresponds to the position of one
brane, and the lower limit might be that of another brane, or else
the position of an event horizon where $\a(r_1)+\Delta(r_1)=0$, if
there is only a single brane.  Comparing to the preceding
discussion, we see that the graviton sees a metric of the form
(\ref{metricform}), with
\eq
    \delta c_g^2 \sim {\int_{r_1}^{r_2} dr\,
    {l \over r}\left[-{\mu\over r^2}+{Q^2\over r^4}
    \right] \over \int_{r_1}^{r_2} dr\, {r\over l} }
\eeq
Interestingly, the sign of $\delta c_g^2$ can be positive or negative,
depending on the relative sizes of black brane mass and charge.

\subsection{Physical Implications}
The physical significance of the metric, $G_{\mu\nu}$, appearing
in the kinetic term of a field within the effective theory, is most
easily seen by working within the geometrical-optics
approximation. Within this approximation, the propagating field is
written in the form $\varphi(x) = A(x) \, \exp[i P_\mu x^\mu]$,
with $A(x)$ assumed to be much more slowly-varying than is the
phase, $P_\mu x^\mu$. With this choice the field equation,
$(G^{\mu\nu} \nabla_\mu \nabla_\nu - m^2)\varphi = 0$ is
equivalent to the dispersion relation: $G^{\mu\nu} \, P_\mu P_\nu
+ m^2 \approx 0$, or equivalently the normal vectors, $P_\mu$, of
the surfaces of constant phase are timelike (or null, if $m=0$)
vectors of the metric $G^{\mu\nu}$.

If the (four-dimensional) wavelength of the mode is much smaller
than the (four-dimensional) radius of curvature of the background
fields, then the motion of these wave packets is along the
geodesics of the 4D-metric $G_{\mu\nu}$. Clearly these
trajectories and dispersion relations generically differ for
fields which have different metrics in their kinetic terms.

The statement is slightly weaker for massless particles, since
the latter move along null geodesics of their respective
metrics. Consequently, their trajectories only differ if the two
metrics are not conformal to one another. In particular,
differences in the trajectories of massless particles are not
observable (in the geometric-optics limit) in the absence of a
preferred frame in the bulk or on the brane.

\section{Loops: General Considerations}
We now turn to the general implications for fermions and photons
of loop-generated Lorentz-violating effects. Motivated by the
considerations of the previous section, we imagine from here on
that all brane fields -- {\it i.e.,} all experimentally observed
elementary particles except for the graviton --  see only the
induced metric on the brane, which we take to be flat and Lorentz
invariant: $\gamma_{\mu\nu} = \eta_{\mu\nu} = \hbox{diag}
(-1,1,1,1)$. By contrast, the metric appearing within the kinetic
terms of any low-energy bulk fields -- which we take to be just
the graviton -- involves the Lorentz-violating metric of
eq.\ \pref{metricform}. This is the dominant low-energy source of
Lorentz violation in the effective theory, and it is the only type
of Lorentz violation whose implications we shall follow.

In general, loops involving virtual bulk states communicate the
news of Lorentz violation to the brane fields, and our task is to
compute the size of this effect. In this section we address
general issues which follow purely from the assumption that
$G^{\mu\nu} = \hbox{diag}(-1/c_g^2,1,1,1)$ encodes all
Lorentz-violating effects, and return in later sections to the
explicit calculation of these effects from graviton loops.

Since we know from direct bounds that Lorentz-violating bulk
effects are small (more about this later), we take $c_g = 1 +
\eps$ with $\eps \ll 1$. Because of the very strong constraints
already known for $c_g < 1$ \cite{MN} our primary interest in what
follows is in positive $\eps$. In view of the direct bounds
arising from solar-system and binary-pulsar tests of general
relativity we imagine $\eps \lsim 10^{-6}$\footnote{A weaker
bound, $\eps \lsim 10^{-3}$, is required if only terrestrial 
bounds, or the gravitational radiation rate of the binary pulsar are used. 
The stronger limit follows from angular momentum conservation 
for the Sun, as inferred by requiring the ecliptic and solar equatorial 
planes not to precess relative to one another throughout the
history of the solar system \cite{Nordtvedt,WillBook,Will}.}.

\subsection{Photon Propagation}
We identify in this section those parts of the graviton-induced
vacuum polarization which have implications for the dispersion
relation of transversely-polarized photons.

We first write the most general form for the vacuum polarization
which can be built from the tensors $\eta_{\mu\nu}$,
$G_{\mu\nu}=\eta_{\mu\nu} + (1-c_g^2) u_\mu u_\nu$ and the
momentum 4-vector, $P_\mu$, which is consistent with symmetry
($\Pi^{\mu\nu} = \Pi^{\nu\mu}$) and transversality ($P_\mu \,
\Pi^{\mu\nu} = 0$). The most general form is:
\eqa \label{GenPi} \Pi^{\mu\nu} &=& A\; \left(\eta^{\mu\nu} -
{P^\mu P^\nu \over P^2} \right) + B \; \Bigl[ u^\mu u^\nu \\
&& \quad \left. + {(P\cdot u)^2 \over P^4} \, P^\mu P^\nu  - \,
{(P \cdot u) \over P^2} \, (P^\mu u^\nu + P^\nu u^\mu) \right],
\nn \eeqa
where at this point $A$ and $B$ are arbitrary functions of the two
independent variables, $P^2$ and $P\cdot u$.

The dispersion relation is found by searching for the zero
eigenvalues of the inverse propagator, $\Delta^{\mu\nu} =
\Delta_0^{\mu\nu} + \Pi^{\mu\nu}$ where $\Delta_0^{\mu\nu} = -
(P^2 \, \eta^{\mu\nu} - P^\mu P^\nu)$. In particular, our interest
is only in those which are transverse (orthogonal to the pure
gauge directions). Working in the rest frame, $u^\mu = (1,0,0,0)$,
and taking the photon momentum to point in the $z$-direction ({\it
i.e.,} $P^\mu = (\omega,0,0,p)$), we therefore require that
$\Delta^{\mu\nu}=0$ in the directions $\mu,\nu = x,y$.

A simple calculation using eq.\ \pref{GenPi} shows that this
implies:
\eq \omega^2 - p^2 + A(\omega,p) = 0. \eeq
Since $A$ is perturbatively small, it suffices to write the
dispersion relation as $\omega = \omega_0 + \omega_1$, where
$\omega_0 = p$, and to evaluate $A$ with $\omega=\omega_0$. This
leads to the present section's main result:
\eq \label{phDispCond} \omega_1 = - \, {1 \over 2 \omega_0} \;
A(\omega_0,p). \eeq

As we shall find, $A$ admits an expansion at low energies in
powers of $P^2$ and $u\cdot P$, leading to the form
\eq A(\omega_0,p) = \alpha \, p^2 + \beta \, p^4 + \dots \, , \eeq
where rotational invariance precludes the appearance of odd powers
of $p$. Using this in eq.\ \pref{phDispCond} and comparing the
result with the general 4D photon dispersion relation
\eq \label{PhForm} \omega^2(p) = p^2 c_\gamma^2 + b_\gamma \,
p^{4} + O(p^6) \, . \eeq
we readily identify
\eq c_\gamma^2 = 1 - \alpha, \qquad b_\gamma = - \beta \, . \eeq

\subsection{Fermion Propagation}
We next ask how loop contributions to fermion self-energies can
modify fermion dispersion relations. We work within the rest frame
defined by $u^\mu$. Writing the inverse electron propagator as $S
= S_0 + \Sigma$, with $S_0 = - ( i \psl + m)$, we see that to
leading order in perturbation theory the zeroes of $S$ satisfy
$P^\mu = (E,\bp)$, where $E= E_0 + E_1$ with $E_0 = \sqrt{p^2 +
m^2_0}$ and:
\eq
\label{DispCond}
i \g^0 E_1 = - \Sigma(E_0,\bp).
\eeq

For small $\bp$, $\Sigma$ has the expansion
\eqa \label{GenSigma}
\Sigma(E_0,\bp) &=& \scA + \scB(i \vec\g
\cdot \bp) +
\scC \, p^2 \nn\\
&& \qquad + \scD\, p^2 (i \vec\g \cdot \bp) + \scE \, p^4 +
\cdots \eeqa
and so eq.\ \pref{DispCond} implies a dispersion relation
of the form
\eq \label{disprel} E_f^2 = m^2_f + p^2\, c_f^2 + b_f \, p^4 +
\cdots \eeq
with
\eqa
m_f &=& m_0 - \scA + \cdots \\
c^2_f &=& 1 - 2(\scB + m_0 \, \scC) + \cdots \\
b_f &=& - 2 (\scD + m_0 \, \scE) + \cdots \eeqa
In these last three equations the subscript $f$ denotes the
fermion species, and the ellipses indicate higher-order
contributions.

Since it is the quantities $c_f^2$ and $b_f$ which we wish to
compare with experiments, the implications of graviton loops may
be obtained by computing the coefficients $\scB$ through $\scE$.

\section{Loops: Graviton Calculations} In this section we compute the
one-loop self energy which is obtained when a fermion or photon
emits and reabsorbs a virtual graviton, as in figs.\ 1 and 2. We
present our results in three steps. First, since the integrals
involved strongly diverge in the ultraviolet, we make some general
remarks about the correct way to treat these divergences before
describing the calculations themselves. We then evaluate the
graviton loop in two steps, motivated by the picture that Lorentz
violation is arising from field configurations within the
extra-dimensional bulk. First we consider loops involving only the
lowest KK mode: the massless 4D graviton. These loops have the
virtue of only involving known particles and couplings, and so the
results we obtain are comparatively robust. They describe the
graviton contributions within the effective 4D theory, well below
the compactification scale, $M_c \sim 1/r$, where $r$ is a measure
of the linear size of the extra dimensions. In the case of a
single noncompact extra dimension, the effective theory is good
below the bulk curvature scale $\sqrt{-\Lambda_5/M_5^3}$, where
$\Lambda_5$ is the bulk cosmological constant and $M_5$ is the 5-D
gravity scale.

Next, we compute the contributions of gravitons in the effective
theory between $M_c$ and the scale $M_l > M_c$ associated with the
extra-dimensional Lorentz violating physics. Since the theory is
extra-dimensional in this energy range, this involves calculating
the loop contributions of higher KK graviton modes. In order to do
this we make several simplifying assumptions about the nature of
the extra-dimensional Lorentz violation, which we believe suffices
for the purposes of establishing the order of magnitude of the
extra-dimensional result.

\subsection{Ultraviolet Divergences}
In $d$ spacetime dimensions the gravitational coupling has
dimension $\kappa_d \sim M_d^{1-d/2}$, where $M_d$ is the
$d$-dimensional Planck mass. On dimensional grounds we therefore
expect the most divergent contribution to one-loop brane-particle
dispersion relations to be
\eq c^2 - 1 \sim \kappa_d^2 \, \Lambda^{d-2}, \qquad b \sim
\kappa_d^2 \, \Lambda^{d-4},  \eeq
where $\Lambda$ is the ultraviolet cutoff scale.

As is usual within an effective theory, this indicates that the
result is most sensitive to the most energetic degrees of freedom
in the problem, suggesting that calculations within the full
theory would produce results that are of order $c^2 -1 \sim
\kappa_d^2 \, M^{d-2}$ and $b \sim \kappa_d^2 \, M^{d-4}$, where
$M$ might be the mass of a heavy particle which was integrated out
to produce the low-energy effective theory.

As we shall see, the above mass-dependence is roughly right,
although some care is required due to the appearance of power-law
divergences \cite{BL}. Care is required because, although the
renormalization group ensures that the coefficient of large
logarithms like $\log(M)$ in observable quantities like $c^2 -1$
or $b$ may be read off from the coefficients of log divergences
(like $\log(\Lambda)$) within the effective theory, the same is
not true for higher (power-law) divergences. As a result in this
section we ignore all power divergences, and compute only the
log-divergent parts of the results. If the results do not have log
divergences (as will be the case with an odd number of extra
dimensions), then we compute only the finite parts of the loops.

Neglecting the power divergences minimizes the Lorentz symmetry
violation which is seen by particles on the brane, and so leads to
conservative conclusions. It corresponds to considering the theory
in which brane-bound particles respect Lorentz invariance at the
energy scale where the theory becomes 4 dimensional. We return in
Section V to the issue of contributions which are proportional to
positive powers of large mass scales, $M$, by considering higher
loops which explicitly involve more massive virtual particles. We
shall there see that naive power-counting estimates do correctly
reproduce the $M$ dependence of the results, but miss important
dimensionless loop factors.

In practice the finite and log-divergent terms are most easily
obtained within dimensional regularization, within which power-law
divergences do not arise. We have computed our results both using
dimensional regularization and using an explicit ultraviolet
cutoff, however, and have verified that the answers obtained are
the same in both cases.

\subsection{Four-dimensional Graviton}
The Feynman rules for fermions and gravitons in the absence of
Lorentz-violating effects are standard, and are obtained by
linearizing the Dirac-Einstein-Hilbert action in curved space,
\eq \label{DEHAction} {\cal L} = - \sqrt{-g} \, \left[ {1 \over 2
\kappa_4^2} \; R + \overline\psi (\dsl + m) \psi + {1 \over 4} \;
F_{\mu\nu} \, F^{\mu\nu} \right], \eeq
about flat space: $g_{\mu\nu} = \eta_{\mu\nu} + 2 \kappa_4 \,
h_{\mu\nu}$. As before $\kappa_4$ denotes the rationalized 
Planck mass in 4D: $M_4 \sim 10^{18}$ GeV. $h_{\mu\nu}$
represents the canonically-normalized graviton field. A recent
statement of the resulting Feynman rules can be found in
references \cite{GRW,HLZ}.

As we have argued in previous sections, we know that the only
Lorentz-violating modification to these rules consists in
replacing the Minkowski metric, $\eta_{\mu\nu}$, with the
nonstandard metric, $G_{\mu\nu}$, when linearizing the first term
in eq.~\pref{DEHAction} to obtain the graviton propagator. In de
Donder gauge this gives:
\eq \label{GravProp} G_{\mu\nu:\alpha\beta}(Q) =
{P_{\mu\nu:\alpha\beta} \over G^{\lambda\rho}\, Q_\lambda Q_\rho -
i\veps} \, , \eeq
where
\eq P_{\mu\nu:\alpha\beta} = \half \left( G_{\mu\alpha}\,
G_{\nu\beta} + G_{\mu\beta} \, G_{\nu\alpha} - G_{\mu\nu} \,
G_{\alpha\beta} \right) ,\eeq
and $\veps$ -- not to be confused with $\eps = c_g-1$ -- is the
infinitesimal which ensures the propagator satisfies Feynman
boundary conditions. The fermion propagators and vertices arise on
the brane, and so use only the usual Minkowski metric.

\subsubsection{The Photon Vacuum Polarization}

\medskip
\centerline{\epsfxsize=2.5in\epsfbox{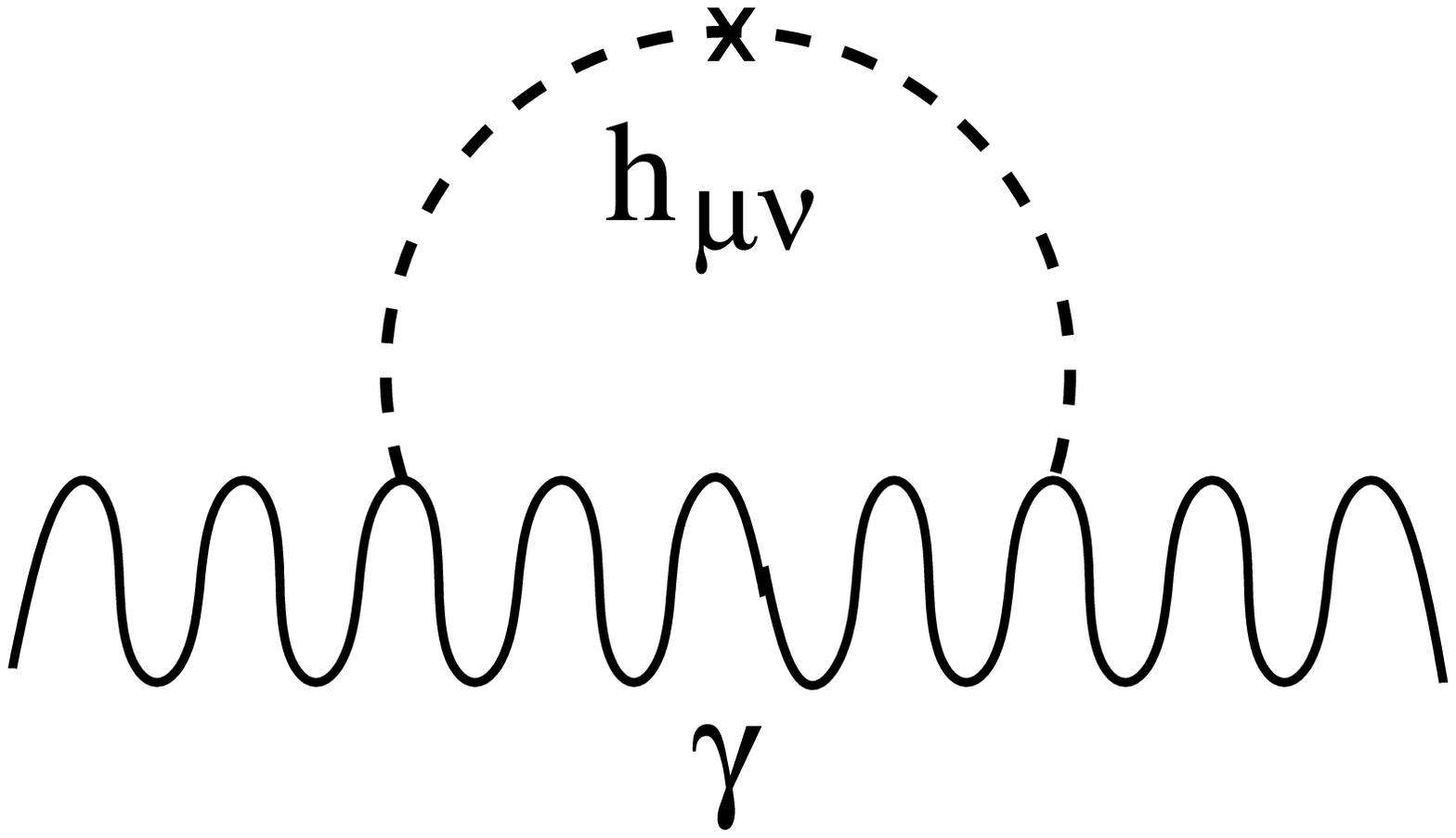}}
{\small
Figure 1. Graviton contribution to photon vacuum polarization.}
\medskip

After Wick rotation, the one-loop Feynman graph in which a virtual
graviton is emitted and reabsorbed by the photon (fig.\ 1) leads
to the following expression for the photon vacuum polarization:
\eq \Pi^{\mu\nu} = \left( {\kappa_4 \over 2} \right)^2 \int {d^4 Q
\over (2\pi)^4} \; {N^{\mu\nu} \over D} , \eeq
where $D = (P-Q)^2 \; G^{\alpha\beta}Q_\alpha Q_\beta$ and
\eq N^{\mu\nu} = V^{\mu\lambda:\alpha\beta}(P,P-Q) \;
{V_\lambda}^{\nu: \sigma\rho}(P-Q,P) \; P_{\alpha\beta:\sigma\rho}
\, . \eeq
Appendix A gives explicit expressions for the vertex functions,
$V^{\alpha\beta:\mu\nu}(P,Q)$. In these expressions all dependence
on the metric $G^{\mu\nu}$ is explicit, and the brane metric,
$\eta_{\mu\nu}$, is to be used to perform any implicit index
contractions, such as in $(P-Q)^2$.

Evaluating this expression (we used the programs FORM and
MATHEMATICA to perform the tensor contractions), Taylor expanding
in powers of $\eps = c_g - 1$ and performing the momentum integral
gives the following expression for the coefficient function, $A$,
of eq.\ \pref{GenPi}:
\eq A(\omega_0, p) = {304 \over 15} \; \lambda_4 \, \eps^2 \, p^4
+ O(\eps^3), \eeq
where $\lambda_4 = (\kappa_4/8\pi)^2 \; {\cal L}$. Here ${\cal L}
= \log(\Lambda^2/\mu^2) = 2/(4-n)$, where the first equality is
evaluated using an ultraviolet cutoff, $\Lambda$, and the second
regularizes by continuing the spacetime dimension, $n$, away from
4. $\mu$ is an arbitrary scale. We ignore the finite part of the
integral relative to its log-divergent part.

Comparison with the general expressions provided earlier gives the
following dispersion coefficients:
\eq \label{phresult}
c_\gamma^2 - 1 = 0, \qquad \qquad b_\gamma =
-\, {304\over 15} \, \lambda_4 \, \eps^2. \eeq

\subsubsection{The Fermion Self-Energy}

\medskip
\centerline{\epsfxsize=2.5in\epsfbox{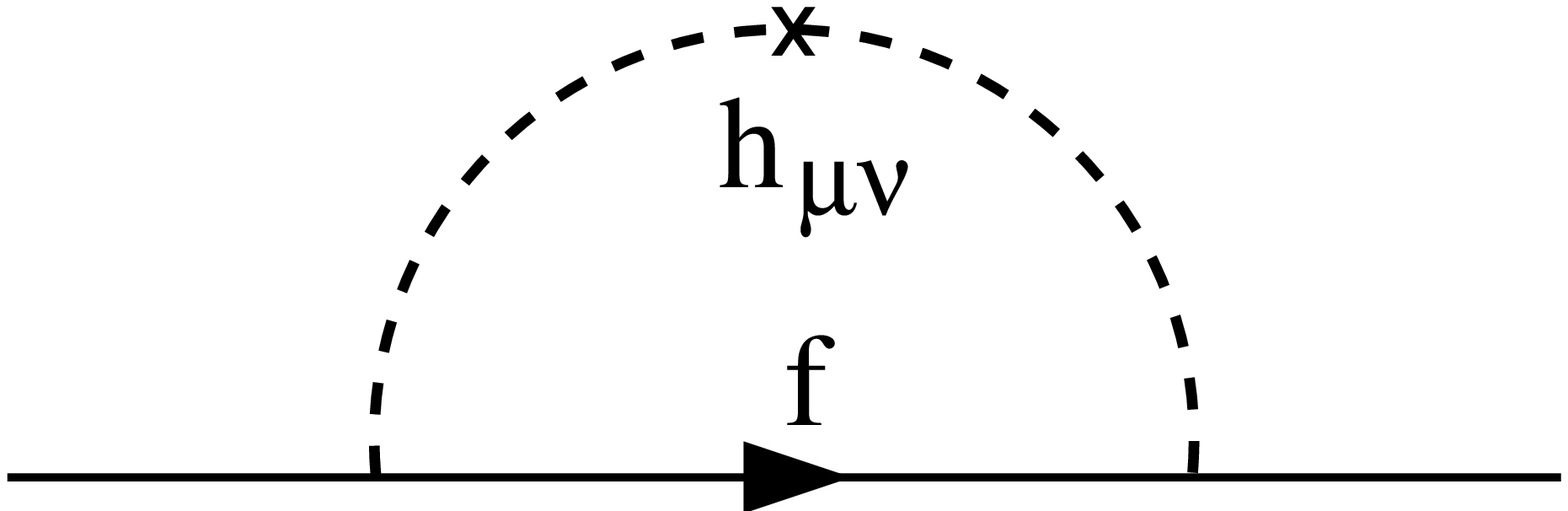}}
{\small
Figure 2. Graviton contribution to fermion self-energy.}
\medskip

We proceed in a similar way for the changes to the fermion self
energy. Evaluating the one-loop Feynman graph using the graviton
propagator of eq.\ \pref{GravProp}, and Wick-rotating to Euclidean
momenta, leads to the following expression for the fermion self
energy:
\eq \Sigma_4 = - \left ( {\kappa_4\over 2} \right)^2 \int {d^4 Q
\over (2\pi)^4} \; {N_1 \over D}, \eeq
where
\eqa \label{None}
N_1 &=& \frac12 \, G_{\mu\nu} \, \gamma^\mu [-i
(\psl - \qsl) + m] \gamma^\nu \nn\\
&&\qquad\qquad \times \; G_{\alpha\beta}(2 P - Q)^\alpha
(2P - Q)^\beta, \nn\\
D &=& G^{\lambda\rho} P_\lambda P_\rho \; [(P-Q)^2 + m^2]. \eeqa
Again all dependence on the metric $G^{\mu\nu}$ is explicit, and
the brane metric, $\eta_{\mu\nu}$, is used to perform the implicit
index contractions in $(P-Q)^2$ and $\psl$. To derive this we used
the simple vertex function $(k_1+k_2)_\mu \gamma_\nu +
(k_1+k_2)_\nu \gamma_\mu$ of ref.\ \cite{GRW} for the
fermion-fermion-graviton coupling rather than the more complicated
one of \cite{HLZ}, which includes an extra term $(i\ksl_1+i\ksl_2
+2m)\eta_{\mu\nu}$. The neglect of this extra term is justified --
even within loop graphs -- because it vanishes if the fermion
field equations are used. This ensures that it is an irrelevant
operator, in the sense that it can be removed by performing a
field redefinition of the fermion of the form $\delta \psi \propto
{h^\mu}_\mu \; \psi$.

We have evaluated this integral to the lowest two orders in the
small parameter $\eps = c_g - 1$, using the programs
FORM/MATHEMATICA to
keep track of the various 4-vectors in the problem. We
find the following results for the logarithmically-divergent part
of the coefficients $\scA$ through $\scE$ of eq.\ \pref{GenSigma}:
\eqa \scA_4 &=&  m^3 \lambda_4 \; \left(4 + 13 \eps + {33\over 2}
\, \eps^2 + ...\right) ,\nn\\
 \scB_4 &=&  m^2 \lambda_4 \; \left(4 \eps + 10 \eps^2 + ...\right) ,\nn\\
 \scC_4 &=&  {m \lambda_4\over 3}  \; \left(16 \eps + 35 \eps^2 + ...\right) ,\nn\\
 \scD_4 &=&   \lambda_4 \; \left(6 \eps^2 + ...\right) ,\nn\\
 \scE_4 &=& 0 + ... \; . \eeqa
where as before $\lambda_4 = (\kappa_4/8\pi)^2 \; {\cal L}$, with
${\cal L} = \log(\Lambda^2/\mu^2) = 2/(4-n)$.

These imply the following contributions to the dispersion relation
of eq.\ \pref{disprel}:
\eqa c_f^2 -1 &=& -{2 m_f^2 \lambda_4 \over 3} \; \left[ 28 \eps +
65 \eps^2 + ... \right], \nn\\
b_f &=& - \lambda_4 \left[ 12 \eps^2 + ... \right] \eeqa
For the case of interest, $\eps > 0$, we see that $c_f^2 < 1$
corresponding to fermions propagating {\it slower} than light. It
follows that (for a given momentum, $p$) the fermion energy is
depressed compared to the photon energy for small $p$.

Notice that the coefficients $b_f$ and $b_\gamma$ both first arise
at $O(\eps^2)$, and so their sign (which is negative for both)
does not depend on the sign of $\eps$. Furthermore, these terms
satisfy $b_f > b_\gamma$. These terms therefore act to raise the
fermion energy relative to the photon, and so act in the opposite
direction of the effect of $c_f < c_\gamma$ (when $\eps
> 0$).

\subsection{Higher-dimensional Gravitons}
We may now estimate the contributions of the higher KK graviton
modes to the propagation of brane-based fermions. Rather than
doing so by performing a sum over a tower of 4-D KK states, we
proceed by directly performing the loop graph using
higher-dimensional gravitons.

The first step is to specify what the higher-dimensional metric is
about which the graviton fluctuation is to be considered. In
principle this should be the metric which describes the
gravitational field of the object or objects in the bulk, whose
presence gives rise to the preferred frame which violates Lorentz
invariance. Since the explicit form for such metrics is rarely
known, we proceed by a more approximate route.

Our approximation is based on the observation that
higher-dimensional graviton loops are ultraviolet sensitive,
with their dominant contributions arising from the circulation of
very short-wavelength modes. So long as the wavelength of these
modes is much shorter than the radii of curvature of the
background Lorentz-violating metric, it should be sufficient to
replace the background metric by one which is approximately flat,
but Lorentz-violating. In particular, this should be sufficiently
accurate for our purposes of estimating the order-of-magnitude of
the resulting loop-generated contributions to fermion propagation
on the brane.

Accordingly we imagine the higher-dimensional graviton to
propagate about a flat metric in which there is a preferred frame
defined by an approximately constant $d$-vector, $u^\mu$. Such a
metric is again described by eq.\ \pref{metricform}, although with
$G_{\mu\nu}$ now being $d$-dimensional.

Linearizing the extra-dimensional Einstein-Hilbert action about
this metric -- $g_{ab} = G_{ab} + 2 \kappa_d \, h_{ab}$ -- we find
the following $d$-dimensional graviton propagator in de Donder
gauge:
\eq \label{GravPropd} G_{\mu\nu:\alpha\beta} = {\half \left(
G_{\mu\alpha}\, G_{\nu\beta} + G_{\mu\beta} \,
G_{\nu\alpha}\right) - {1\over d-2} \; G_{\mu\nu} \,
G_{\alpha\beta} \over G^{\lambda\rho} P_\lambda P_\rho -i\veps} \,
. \eeq
Again $\veps$ in this expression is the infinitesimal which
enforces Feynman boundary conditions. The quantity $\kappa_d$ is
related to the $d$-dimensional Newton's constant and Planck mass
by $\kappa_d^2 = 8 \pi G_d = (1/M_d)^{d-2}$. We take the fermions
and photons to move on a 3-brane, for which the induced metric is
the usual Minkowski metric.

\subsubsection{The Photon Vacuum Polarization}
Computing the photon vacuum polarization with this propagator
gives no correction to the photon dispersion relation at low
energies, if the number of dimensions exceeds the usual four.
This can be understood purely in terms of dimensional analysis,
and the fact that we are computing only the finite or log-divergent
contributions. The one-loop contributions to the photon
self-energy have two graviton vertices, so they are proportional
to $\kappa^2_d = M_d^{-(d-2)}$. Since the log-divergent and finite parts do not
involve powers of the cutoff $\Lambda$, the only quantity which
can be used to make a dimensionally correct answer is the photon
momentum, $p$. Thus the result is proportional to $\kappa_d^2
\, p^{d}$.  This gives a $p^4$ correction to the dispersion relation,
but only if $d=4$. For $d>4$ the correction always gives only a higher
than quartic power of $p$. (Section V discusses the physics of the
power-divergent contributions to $c_\gamma^2 - 1$ and $b_\gamma$.)

\subsubsection{The Fermion Self-Energy}
Evaluating the one-loop fermion self-energy graph using the
$d$-dimensional graviton propagator of eq.\ \pref{GravPropd}, and
Wick-rotating to Euclidean momenta, leads to the following
expression for the fermion self energy:
\eq \Sigma_d = - \left ( {\kappa_d\over 2} \right)^2 \int {d^d Q
\over (2\pi)^d} \; \left( {N_1 + N_2 \over D} \right), \eeq
where $N_1$ is as given by eq.\ \pref{None} and
\eqa N_2 &=& C_d \, \Bigl\{ [i(\psl {-} \qsl) {+} m ]
G_{\mu\nu} \, (2 P {-} Q)^\mu (2 P {-} Q)^\nu \nn\\
&& \; \, -2i G_{\alpha\beta} \gamma^\alpha (2 P {-} Q)^\beta
G_{\lambda\rho} (P {-} Q)^\lambda (2P {-} Q)^\rho \Bigr\},\eeqa
with $C_d =  (d-4)/[2(d-2)]$.

Evaluating the integral in powers of $\eps = c_g - 1$ gives an
ultraviolet divergent result. We have evaluated the result using
both dimensional regularization and an explicit cutoff, and have
verified that the coefficients of the logarithmically-divergent
and pole terms are the same in both cases. For odd dimensions the
integrals are finite in dimensional regularization, and we have
verified that the result agrees with the finite part when
evaluated with an ultraviolet cutoff.

We are led in this way to the following expressions for the
quantity $c_f^2 - 1$ for dimensions $d=5$ through 10. (We give the
quantities $\scA$ through $\scE$ in Appendix B.):
\eqa c_f^2 {-}1 &=&  \!\phm m_f^3 \, \lambda_5  \: \left[ \frac{110}{9} \,
\eps + \frac{239}{9} \, \eps^2 + ... \right],  \quad (d=5)\nn\\
&=&\!  \phm m_f^4 \, \lambda_6  \: \left[ \frac{\: 48\: }{5}
\, \eps + \frac{\; 99 \;}{5} \, \eps^2 + ... \right],  \quad (d=6)\nn\\
&=&\! -\, m_f^5 \, \lambda_7  \: \left[ \frac{616}{75}
\, \eps + \frac{244}{15} \, \eps^2 + ... \right],   \quad (d=7) \nn\\
&=&\! -\, m_f^6 \, \lambda_8  \: \left[ \frac{464}{63}
\, \eps + \frac{890}{63} \, \eps^2 + ... \right],  \quad (d=8)
\nn\\
&=&\!  \phm m_f^7 \, \lambda_9  \: \left[ \frac{333}{49}
\, \eps + \frac{\!1245\!}{98} \, \eps^2 \!+ ... \right],  \quad (d=9) \nn\\
&=&\! \phm m_f^8 \, \lambda_{10\!} \left[ \frac{115}{18} \, \eps +
\frac{421}{36} \, \eps^2 + ... \right], \quad (d=10). \! \eeqa
Here the quantities $\lambda_d$ are defined in terms of the
couplings $\kappa_d$ by:
\eqa \lambda_5 = {\kappa_5^2\over (8 \pi)^2}, \phl \quad \;\;
\lambda_6 &=& {2\kappa_6^2\over (8 \pi)^3} \; {\cal L}, \quad \;\;
\lambda_7 \,= {\kappa_7^2 \over 6(4\pi)^3}, \\
\lambda_8 = {2\kappa_8^2 \over (8 \pi)^4} \; {\cal L}, \quad \;\;
\lambda_9 &=& {\kappa_9^2\over 15 (4 \pi)^4}, \quad \;\;
\lambda_{10\!\!} = {4\kappa_{10}^2 \over 3(8 \pi)^5} \; {\cal L}.
\nn \eeqa
As before ${\cal L} = \log(\Lambda^2/\mu^2)$ when a cutoff is
used, or ${\cal L} = 2/(d-n)$ in dimensional continuation of $n$
away from $n=d$.

Notice that the corrections to $c_f^2$ which are implied in this
way are not universal in size for all fermions, being suppressed
by powers of $m_f$ for lighter fermions.

The corresponding higher-order dispersion coefficient, $b_f$, is
similarly:
\eqa b_f &=&   \phm m_f \, \lambda_5 \,\left[ \frac{26}{3} \, \eps^2 +
... \right]  , \qquad (d=5) \nn\\
&=&   \phm m_f^2 \, \lambda_6 \,\left[ \frac{36}{5} \, \eps^2 + ...
\right] , \qquad (d=6) \nn\\
&=& -\, m_f^3 \, \lambda_7 \,\left[ \frac{32}{5} \, \eps^2 + ... \right],
\qquad (d=7) \nn\\
&=& -\, m_f^4 \, \lambda_8 \,\left[ \frac{\! 124 \!}{21} \, \eps^2 \!+ ...
\right] , \qquad (d=8) \nn\\
&=&  \phm m_f^5 \, \lambda_9 \,\left[ \frac{39}{7} \, \eps^2 + ...
\right] , \qquad (d=9) \nn\\
\label{bfeqs}
&=&  \phm m_f^6 \, \lambda_{10\!\!} \left[ \frac{16}{3} \, \eps^2 + ...
\right] , \qquad (d=10) . \eeqa
Just as we saw for $d=4$, $b_f$ always arises at second order in
$\eps$, and so its sign is completely determined in our
calculation. As we shall see, the best bound on this coefficient
arises when $b_f > b_\gamma \approx 0$.

The above expressions were derived specifically with the scenario
of large, flat extra dimensions in mind.  However we can also
interpret at least the 5D result in terms of a warped extra
dimension.  This has more interesting consequences than the flat
case, where the quantum gravity scale would have to exceed
$M_5\gsim 10^8$ GeV in order to  comply with sub-millimeter tests
of gravity \cite{adelberger}.  In the warped case, even if
$M_5\sim M_p$, the KK gravitons couple to the  TeV brane with TeV
strength, and they have a mass gap of order TeV.   In this model,
we live on a ``TeV brane'' at $r=r_1$, displaced from the ``Planck
brane'' at $r=r_2$ such that ${r_1\over r_2}\sim 10^{-16}$ in
accordance with solving the hierarchy problem.  As far as the
contributions from the ultraviolet graviton loops are concerned,
this looks like quantum gravity with a scale of $M_5\sim$ TeV. The
TeV mass gap protects low energy gravity from any observable
distortion, but not so the loop effects from momenta $p\gg$ TeV.

\section{Real-World Complications}
A further step is required before these results can be compared
with the experimental limits on the dispersion relations of real
particles. This step involves identifying which low-energy
particles produce the largest contributions to any given
dispersion relation.\footnote{We thank M.~Pospelov for making many
very useful suggestions for this and the next sections.}

There are two reasons why this additional step is required. First,
we would like to apply the above calculations to protons, for
which the experimental limits are the strongest. Unfortunately,
our calculations treat all fermions as elementary, and so can only
apply directly to the proton for scales below roughly $m_p \sim 1$
GeV. (Notice these low energies may nonetheless be described by an
extra-dimensional effective theory within ADD-type scenarios.) For
higher energies, we must apply our calculations to the constituent
quarks and gluons, and infer from these how the proton dispersion
is affected. Unfortunately, the resulting strong-interaction
uncertainties prohibit us from following in more detail the $O(1)$
factors and signs of the results, limiting us to an
order-of-magnitude analysis for the proton.

The second reason for being careful in applying our results is the
strong dependence which they have on the relevant fermion masses. In
particular, it may be that larger contributions are obtained for light
particles (like electrons and photons) by embedding the
graviton-induced Lorentz-violating contributions of heavier particles
(like top quarks) within additional loops. This is how we will recover
the larger contributions that might have naively been obtained by
keeping the power divergences of the graviton loop graphs.

\subsection{Photons}
Our result for the photon dispersion relation is particularly
simple, with $c_\gamma = 1$ for all $d$, and $b_\gamma \ne 0$ only
for $d=4$. This simplicity is largely due to our concentrating on
the finite and log-divergent contributions, however, which
suggests that larger contributions may be found at the
(comparatively cheap) expense of introducing additional loops.

For instance, at two loops the photon vacuum polarization acquires
contributions by inserting a Lorentz-violating graviton
self-energy within a charged-fermion loop (fig.\ 3). Assuming all
charged fermions to be brane-bound, and so 4-dimensional, we
estimate the following results for the photon: $c_\gamma^2 -1 \sim
\left( {\alpha \over 4 \pi} \right) \, (c^2_f - 1)$, giving
\eqa \label{phest}
c_\gamma^2 -1 &\sim& \left( {\alpha \over 4 \pi}
\right) \left( {\eps \over (4 \pi)^{[d/2]}} \right) \; \left( {m_f
\over
M_d} \right)^{d-2} \\
&\sim& \left\{ \matrix{ 1\times 10^{-10} \, \left( {\eps \times
10^{3}} \right) \, \left( { \hbox{TeV} / M_4} \right)^2 & (d=4)\cr
2 \times 10^{-11} \, \left( {\eps \times 10^{3}} \right) \, \left(
{ \hbox{TeV} / M_5} \right)^3 & (d=5) \cr
3 \times 10^{-13} \, \left( {\eps \times 10^{3}} \right) \, \left(
{ \hbox{TeV} / M_6} \right)^4 & (d=6) \cr} \right.  \, ,\nn \eeqa
where $[d/2]$ denotes the integer part of $d/2$ and we use the
heaviest known elementary particle, the top quark ($m_t = 175$
GeV), for numerical purposes. Although this is still negligibly
small for $d=4$, we shall see that the $d>4$ result can be large
enough to provide new constraints on $\eps$ if $M_d$ is not too
far above the TeV range.

A similar contribution arises to $b_\gamma$, although because of
the weaker dependence on $m_f$ the price of a loop factor,
$\alpha/4\pi$, is not worthwhile when $d=4$, where the direct
one-loop result of eq.~\pref{phresult} is larger. We expect,
then
\eqa \label{phestbd4} b_\gamma M_4^2 &=& - \, {304\over 15} \left(
{\eps \over 8 \pi} \right)^2 \qquad (d = 4)\\
&=& -3 \times 10^{-8} \, \left( {\eps \times 10^{3}}
\right)^2\eeqa
in four dimensions, where the numerical result uses the
conservative estimate $\log(\Lambda^2/\mu^2) \sim 1$.

For $d>4$ we estimate the two-loop result by $b_\gamma \sim \left(
{\alpha \over 4 \pi} \right) \, b_f$, and so find:
\eqa \label{phestb} b_\gamma M_d^2 &\sim& \left( {\alpha \over 4
\pi} \right) \left( {\eps^2 \over (4 \pi)^{[d/2]} }
\right) \; \left( {m_f \over M_d} \right)^{d-4} \qquad (d>4)\\
&\sim& \left\{ \matrix{ 7 \times 10^{-13} \, \left( {\eps \times
10^{3}} \right)^2 \, \left( { \hbox{TeV} / M_5} \right) & (d=5)
\cr 1 \times 10^{-14} \, \left( {\eps \times 10^{3}} \right)^2 \,
\left( { \hbox{TeV} / M_6} \right)^2 & (d=6) \cr} \right.  \, ,\nn
\eeqa
with $m_t$ being used for the fermion mass. The sign of $b_\gamma$
and $c_\gamma^2 -1$ are not determined by these estimates
(although they are certainly calculable within a more careful
evaluation of Fig.~(3).

\medskip
\centerline{\epsfxsize=2.5in\epsfbox{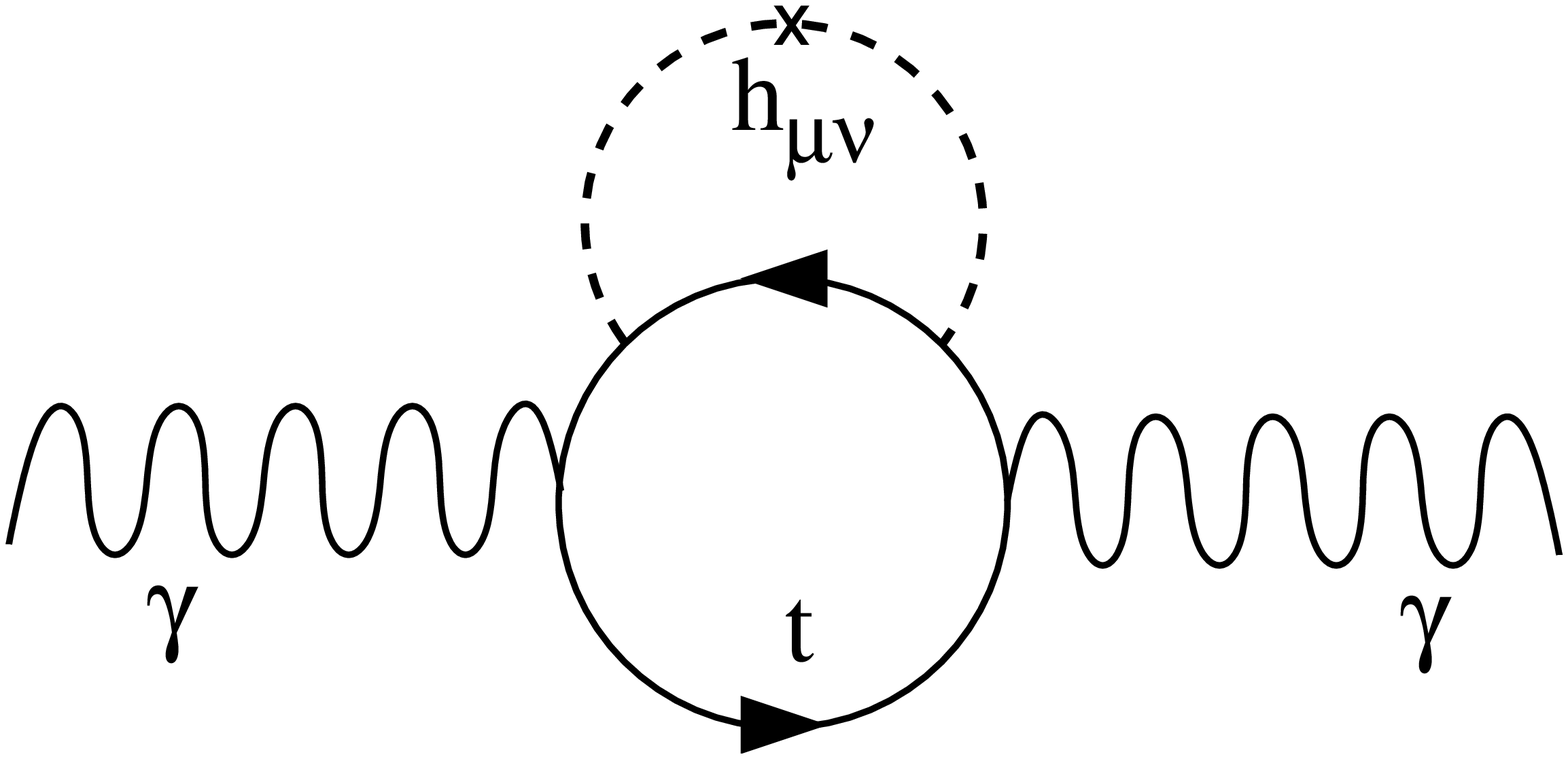}}
{\small
Figure 3. Two loop contribution to photon vacuum polarization from
top quark and graviton.}

\subsection{Electrons}
The next-cleanest application of the above analysis is to
electrons, since so far as we know these are elementary and so our
earlier results may be directly applied. Again taking the
conservative estimate $\log(\Lambda^2/\mu^2) \sim 1$, we find that
the direct graviton-loop contribution to the electron dispersion
relation is
\eqa \label{eldir} c_e^2 - 1 &\sim& { \eps \over (4 \pi)^{[d/2]}}
\; \left( {m_e
\over M_d} \right)^{d-2}, \nn\\
b_e &\sim& {\eps^2 \over (4 \pi)^{[d/2]} \, M_d^2} \; \left( {m_e
\over M_d} \right)^{d-4}, \eeqa
where $d$ counts the number of spacetime dimensions seen by the
graviton within the effective theory at scales $\mu < \Lambda \sim
1$ TeV.

This should be compared with the result of inserting more massive
particles into higher loops. For instance, a loop with a W boson and
neutrino, with the graviton coupling to the W (fig.\ 4)
gives the alternative contribution
\eqa \label{elest} c_e^2 -1 &\sim& \left( {\alpha_w \over 4 \pi}
\right) \left( {\eps \over (4 \pi)^{[d/2]}} \right) \; \left(
{m_w \over M_d} \right)^{d-2} \\
&\sim& \left\{ \matrix{ 1\times 10^{-10} \, \left( {\eps \times
10^{3}} \right) \, \left( { \hbox{TeV} / M_4} \right)^2 & (d=4)\cr
3 \times 10^{-12} \, \left( {\eps \times 10^{3}} \right) \, \left(
{ \hbox{TeV} / M_5} \right)^3 & (d=5) \cr
6 \times 10^{-14} \, \left( {\eps \times 10^{3}} \right) \, \left(
{ \hbox{TeV} / M_6} \right)^4 & (d=6) \cr} \right.  \, ,\nn \eeqa
using $m_w = 80$ GeV, and $\log(\Lambda^2/mu^2) = 1$.
This last contribution (for $d \ge 4$) is always larger than the
direct one, eq.\ \pref{eldir}, because $(m_e/m_w)^{d-2} \sim (6
\times 10^{-6})^{d-2} \ll (\alpha_w /4 \pi) \sim 3\times 10^{-3}$.

\medskip
\centerline{\epsfxsize=2.5in\epsfbox{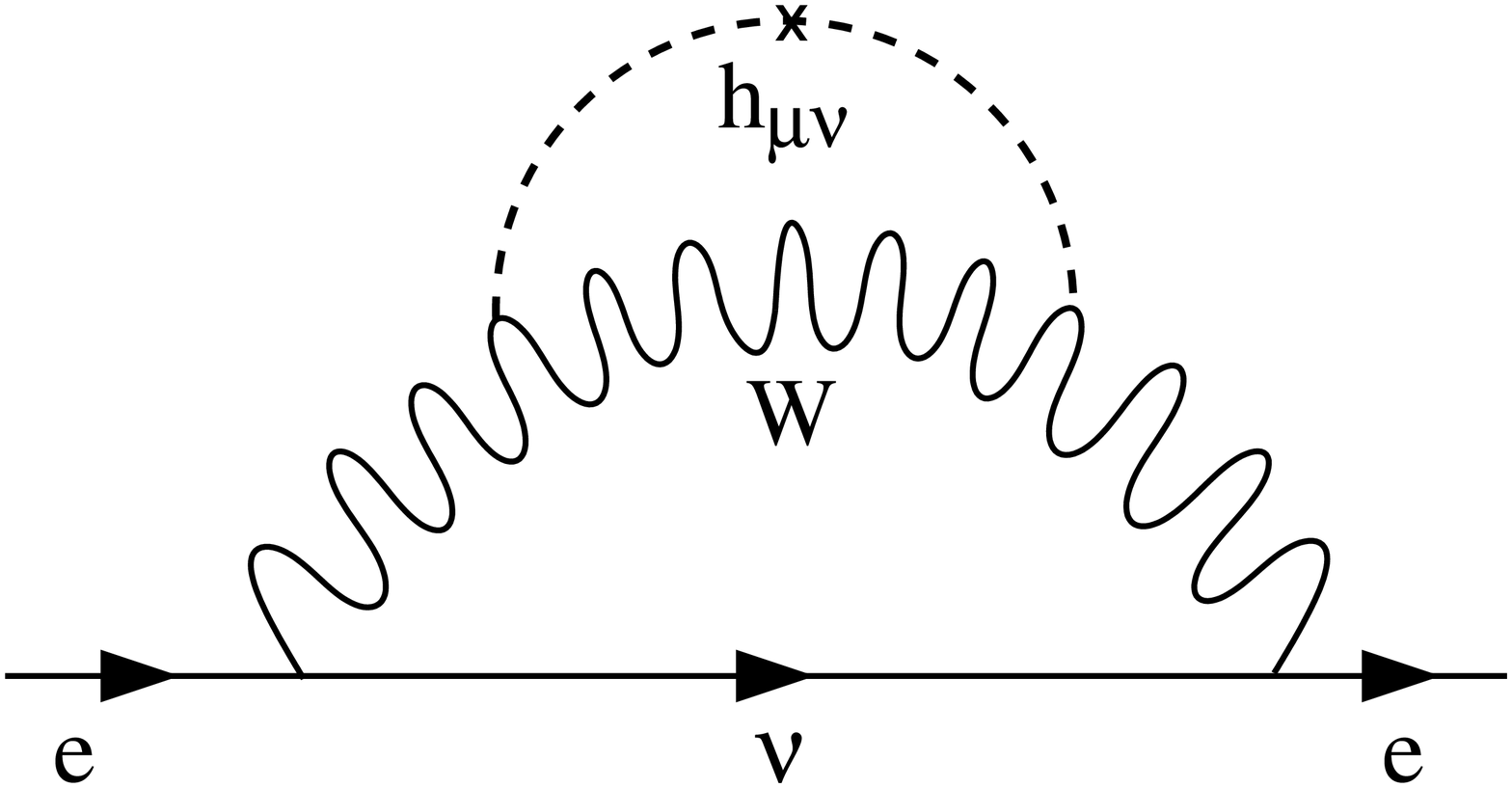}} {\small Figure
4. Two-loop contribution to electron self-energy from W boson
and graviton, which can dominate over the one-loop graph
with gravitons alone, Fig.\ 2.}
\medskip

The higher-loop result can also provide a larger estimate for
$b_e$, to wit:
\eqa \label{elestb} b_e M_d^2 &\sim& \left( {\alpha_w \over 4 \pi}
\right) \left( {\eps^2 \over (4 \pi)^{[d/2]} }
\right) \left( {m_w \over M_d} \right)^{d-4} \qquad (d\ge 5)\\
&\sim& \left\{ \matrix{ 4 \times 10^{-13} \, \left( {\eps \times
10^{3}} \right)^2 \, \left( { \hbox{TeV} / M_5} \right) & (d=5)
\cr 9 \times 10^{-15} \, \left( {\eps \times 10^{3}} \right)^2 \,
\left( { \hbox{TeV} / M_6} \right)^2 & (d=6) \cr} \right.  \, .\nn
\eeqa
This dominates the direct one-loop result for all $d \ge 5$,
because $(m_e/m_w)^{d-4} \sim (6 \times 10^{-6})^{d-4} \ll
(\alpha_w /4 \pi) \sim 3\times 10^{-3}$.

When $d=4$ the direct one-loop result wins, for which the results
of the previous section give
\eqa \label{elestbd4} b_e M_4^2 &=& - 12 \left( { \eps \over 8
\pi } \right)^2 \qquad (d= 4) \\
&=& -2 \times 10^{-8} \, \left( {\eps \times 10^3} \right)^2 .\nn\\
\eeqa

\subsection{Protons}
The direct calculation of section III applies directly for the
proton only within the effective theory below a GeV, because it is
only within this theory that the proton may be considered to be
elementary. For higher scales the relevant degrees of freedom are
quarks and gluons, for which we must estimate the size of
Lorentz-violating effects.

The Lorentz-violating gluon and quark contributions may be estimated using our
photon and electron results, eqs.~\pref{phest} and \pref{elest}, giving:
\eqa \label{glest}
c_g^2 -1 &\sim& \left( {\alpha_s \over 4 \pi}
\right) \left( {\eps \over (4 \pi)^{[d/2]}} \right) \; \left( {m_t
\over M_d} \right)^{d-2} \nn\\
c_q^2 -1 &\sim& \left( {\alpha_w \over 4 \pi}
\right) \left( {\eps \over (4 \pi)^{[d/2]}} \right) \; \left(
{m_w \over M_d} \right)^{d-2} \eeqa
where $\alpha_s$ is the QCD coupling, which we take to be $0.119$.
From this we estimate the
proton result to be
\eqa \label{prest}
c_p^2 -1 &\sim& \hbox{Max}(c_g^2 -1,c_q^2 -1) \nn\\
&\sim& \left( {\alpha_s \over 4 \pi}
\right) \left( {\eps \over (4 \pi)^{[d/2]}} \right) \; \left( {m_t
\over M_d} \right)^{d-2} \nn\\
&\sim& \left\{ \matrix{ 2\times 10^{-9} \, \left( {\eps \times
10^{3}} \right) \, \left( { \hbox{TeV} / M_4} \right)^2 & (d=4)\cr
3 \times 10^{-10} \, \left( {\eps \times 10^{3}} \right) \, \left(
{ \hbox{TeV} / M_5} \right)^3 & (d=5) \cr
4 \times 10^{-12} \, \left( {\eps \times 10^{3}} \right) \, \left(
{ \hbox{TeV} / M_6} \right)^4 & (d=6) .\cr} \right. \nn \eeqa%

The loop contributions to $b_p$ can also dominate for large
numbers of dimensions, but not for $d=4$. We have the higher-loop
contribution
\eqa \label{prestb} b_p M_d^2 &\sim& \left( {\alpha_s \over 4 \pi}
\right) \left( {\eps^2 \over (4 \pi)^{[d/2]} }
\right) \; \left( {m_f \over M_d} \right)^{d-4} \qquad (d\ge 5)\\
&\sim& \left\{ \matrix{ 1 \times 10^{-11} \, \left( {\eps \times
10^{3}} \right)^2 \, \left( { \hbox{TeV} / M_5} \right) & (d=5)
\cr 1 \times 10^{-13} \, \left( {\eps \times 10^{3}} \right)^2 \,
\left( { \hbox{TeV} / M_6} \right)^2 & (d=6) \cr} \right.  \, ,\nn
\eeqa
when $m_f = m_t$.

By contrast, for $d=4$ it is the direct one-loop result which
dominates. In this case we have a result for fermions which is
largely insensitive to the fermion mass:
\eqa \label{prestbd4} b_f M_4^2 &=& - 12 \left( { \eps \over 8
\pi } \right)^2 \qquad (d= 4) \\
&=& -2 \times 10^{-8} \, \left( {\eps \times 10^3} \right)^2 \, .
\nn \eeqa
For low-energy gravitons in the effective theory below the proton
mass, the proton may be considered to be elementary and
eq.~\pref{prestbd4} can be directly applied to the proton itself.

For higher-energy gravitons, eq.~\pref{prestbd4} would instead be
applied to the light quarks, and the result for the proton would
be obtained by taking the matrix element for the resulting
effective quark operator, such as ${\cal O}_{\rm eff} = \eps^2 \,
u^\mu u^\nu u^\alpha u^\beta \; \overline{q} \gamma_\mu
\partial_\nu \partial_\alpha \partial_\beta \, q$, within the
proton. On dimensional grounds this would produce the same result
as eq.~\pref{prestbd4}, but potentially with a different numerical
coefficient. For concreteness, when comparing with the observables
we use eq.~\pref{prestbd4} including the numerical factor obtained
for an elementary proton.

Notice that the above estimates suggest that so long as
higher-loop contributions dominate, a hierarchy is to be expected:
$| c_p^2 -1 | \gg | c_\gamma^2 - 1 | \sim | c_e^2 - 1|$. The same
would {\it not} be expected to apply for $|b_p|, |b_\gamma|$ and
$|b_e|$ when $d=4$, however.

\section{Experimental Constraints}
We now turn to the experimental constraints which can be imposed
on the quantities $c^2 -1$ and $b$. Because we are interested in
order-of-magnitude bounds, we derive constraints on each of these
quantities as if they had arisen in the absence of the other. That
is, we consider bounds on $c^2 -1$ while taking $b =0$ and vice
versa. This is justified unless there is an unnatural cancellation
between the contributions of these quantities to physical
observables.

\subsection{Existing Bounds on $c_f^2 - c_\gamma^2$}
Good bounds exist on a difference between the propagation speeds
of fermions and photons. Among those which do not depend on the
sign of $c_f -c_\gamma$ are \cite{Glashow}:
\eqa |c_f - c_\gamma | &<& 6 \times 10^{-22} \qquad \hbox{Atomic
spectroscopy} \nn\\
|c' - c|_{\mu e} &<& 4 \times 10^{-21} \qquad \mu \to e\gamma
\nn\\
|c_{\sss KL} - c_{\sss KS}| &<& 3 \times 10^{-21} \qquad K-\overline{K}
\quad\hbox{oscillations} . \eeqa
In the second bound $c'$ denotes a particular combination of the
Lorentz-violating couplings in flavor space, whose details are
not important in what follows.

Although these bounds apply to both signs of $c^2 -c_\gamma^2$,
they are also subject to specific assumptions which need not apply
to our calculation. For instance the $\mu\to e\gamma$ bound
assumes the existence of Lorentz-violating terms which also
violate lepton number, while the bounds involving neutral kaons
require strangeness violation in addition to Lorentz-violation.
Since the loops which would produce both type of flavor-symmetry
violations from Lorentz-violation in the graviton sector are
further suppressed by masses and mixing angles, we do not consider
these bounds further.

The bounds from atomic spectroscopy are usually derived under the
assumption that all matter particles share the same maximum
propagation speed which differs from that of the photon -- {\it
i.e.,} $c_f \ne c_\gamma$ is independent of $f$. Nevertheless
these bounds are likely also to apply if $c_f$ differs for
electrons and nucleons. Although we use the bound of
ref.~\cite{Glashow} in what follows, we regard a more careful
analysis of constraints which are implied by these experiments to
be worth pursuing.

If we apply these bounds directly to protons, we find $|c_p^2 -
c_\gamma^2| < 6 \times 10^{-22}$. Using our previous estimates for
the proton, eq.~\pref{prest}, and our expectation (see above) that
$c_p^2$ differs from unity by more than does $c_\gamma^2$, we find
comparison with the bound implies,
\eq \eps < \left( {4 \pi \over \alpha_s} \right)
\; (4 \pi)^{[d/2]} \, \left( {M_d \over m_t} \right)^{d-2}
10^{-21}. \eeq

This is stronger than the direct bound $\eps
<10^{-6}$\footnote{Using the weaker bound $\eps < 10^{-3}$ makes
the limits for $M_d$ larger by a factor of $10^{3/(d-2)}$.} only
if
\eq {M_d \over \hbox{TeV}} < 0.2 \; \left[ {10^{13} \over (4
\pi)^{[d/2]}} \right]^{1/(d-2)} \sim \left\{ \matrix{3\times 10^4
& (d=4) \cr 700 & (d=5) \cr 46 & (d=6) } \right. .\eeq
The bound so obtained is not useful for $d=4$. With extra
dimensional gravitons the bound becomes useful only when $M_d$ is
very small. It is only of borderline interest for the ADD
scenario, for which $d=6$ but $M_6 > 50$ TeV is required from
stellar-cooling bounds \cite{ADDBounds}. (We do not quote here the
stronger limits coming from the non-observance of gamma-ray decay
products \cite{ADDGBounds}, or which apply for supersymmetric
models \cite{LED}, because these may be evaded depending on
model-dependent details.)

The bound {\it is} competitive, however, for the warped 5D model
described below eq.\ (\ref{bfeqs}), where $M_5$ is effectively the
TeV scale.

\subsection{\v Cerenkov Bounds on $b_f$ and $c_f^2 -1$}

We next consider constraints from high-energy cosmic rays,
typically protons or photons.  Their observation precludes the existence of
processes which would too-efficiently deplete the energy of these
particles. Since the particles involved are at higher energies, these
processes are more sensitive to the higher-momentum coefficients,
$b_f, b_\gamma$, than are low-energy laboratory limits.

Lorentz-violation can introduce dangerous processes by allowing decays
in vacuum of the form $p \to p\gamma$ or $\gamma \to e^+ e^-$ or
$\gamma \to p \overline{p}$. Such decays are precluded by energy
and momentum conservation in Lorentz-invariant systems, but {\it
are} allowed given Lorentz-violating dispersion relations. For
instance, the process $p \to p \gamma$ becomes allowed if the
dispersion relation raises the energy of the proton at a given
momentum more than it does for the photon. Similarly, the process
$\gamma \to e^+ e^-$ can occur if the photon energy is raised more
than the electron's at a given momentum.

As applied to changes in the maximum propagation speed, the
resulting bounds are therefore one-sided, in the sense that they
are only relevant for one sign of $c_p - c_\gamma$ or $c_\gamma -
c_e$. For protons the strong one-sided bound obtained from $p \to
p\gamma$ requires $c_p > c_\gamma$. Similarly, for $\gamma \to e^+
e^-$ or $\gamma \to p \overline{p}$, $c_e < c_\gamma$ and $c_p <
c_\gamma$ are required to use the one-sided bound.

Keeping in mind that our estimates of the previous sections imply
$|\delta c_p^2| \gg |\delta c_\gamma^2 | \sim | \delta c_e^2 |$, we
find the limits obtained in this way for electrons and protons are
\eqa -2 \times 10^{-8} < c_p^2 - c_\gamma^2 \sim c_p^2 -1 &<& 2 \times 10^{-23} \nn\\
c_e^2 - c_\gamma^2 &>& -6 \times 10^{-15}. \eeqa
In order of magnitude, these bounds are obtained by demanding that
$|c_f^2 -c_\gamma^2 |< m_f^2/E^2$ where $m_f$ is the relevant
fermion mass, $E$ is the energy of the observed cosmic ray, and
the bound only applies to the appropriate sign of $c_f^2 -
c_\gamma^2$. The bound from $\gamma \to f \overline{f}$ is weaker
than that from $p \to p \gamma$ because the most energetic cosmic
ray proton has $E \sim 10^{8}$ TeV, while the most energetic
observed gamma ray has $E \sim 50$ TeV.

A similar bound also applies to the parameter $b_f$, which
controls the dispersive part of the dispersion relation, eq.\
\pref{disprel}. For $p \to p\gamma$ the bound then applies only if
$b_p > b_\gamma$ and for $\gamma \to f \overline{f}$ it requires
$b_f < b_\gamma$. When the bound applies, it is of order $|b_f -
b_\gamma | < m_f^2/E^4$.

We find in this way the constraints
\eqa -(5 \times 10^{8} \; \hbox{GeV})^{-2} < &b_p - b_\gamma& < (3
\times 10^{22} \;
\hbox{GeV})^{-2} \nn\\
&b_e - b_\gamma & > - (1 \times 10^{12} \; \hbox{GeV})^{-2} .
\eeqa

\subsubsection{$b_p > b_\gamma$}
Consider first the case $b_p > b_\gamma$, for which the best bound
applies. For $d>4$ we compare this to the order-of-magnitude of
the higher-loop results, with the resulting expectation $|b_p| \gg
|b_\gamma|$. Using eq.~\pref{prestb} for $b_p$, as discussed in
the previous section, and assuming the sign of the result is the
one relevant for the bound's applicability, we obtain the limit
\eq \eps < \left[ {(4 \pi)^{[(d+2)/2]}\over{\alpha_s}} \, \left( {M_d \over \ m_t}
\right)^{(d-4)} \right]^{1/2} \; \left( { M_d \over 3 \times
10^{22} \; \hbox{GeV} } \right). \eeq

This is an improvement over $\eps < 10^{-6}$ if
\eqa {M_d \over \hbox{TeV}} &<& \left[ { 3 \times 10^{13} \,  \ (
{0.175})^{(d-4)/2} \sqrt{{\alpha_s}\over (4 \pi)^{[d/2]+1}}}
\right]^{2/(d-2)} \nn\\
&<& \left\{ \matrix{ 2 \times 10^{7}&(d=5) \cr 9 \times
10^4&(d=6)\cr 6 \times 10^4 & (d=7) \cr 800 &(d=8)\cr 300
&(d=9)\cr 70&(d=10) \, .} \right. \eeqa
(An improvement on the bound $\eps < 10^{-3}$ is obtained for
$M_d$ which can be a factor $10^{6/(d-2)}$ larger.) We see the
bound can be competitive for all $d$, provided $M_d$ is in the
lower end of its allowed range. It is particularly strong for the
ADD ($d=6$) and warped RS ($d=5$) cases, for which $M_d$ is in the
TeV range.

To apply the bound to $d=4$ we instead use the one-loop result,
for which the numerical factors and sign are known from our
previous calculation (provided that the high-energy quark
contribution does not dominate that for the low-energy proton, as
discussed above). Although $b_p$ which is obtained is negative,
the bound nonetheless applies because for $d=4$ we know that
$b_\gamma$ is also negative, with $b_f > b_\gamma$ ({\it c.f.}
eqs.~\pref{phestbd4} and \pref{prestbd4}).%
\footnote{
In fact, at the energies under consideration the graviton should resolve
the proton as a number of partons, each carrying a small momentum
fraction $x \ll 1$.  Since the importance of the
dispersive term grows with momentum as
$p^2$, there will be a suppression $\sim x_{\rm av}^2$.  Hence,
compared to the photon, the effective $b$ for the proton may be close to
zero.}
We find
\eq \label{d4bound} b_p - b_\gamma = 10^{-8} \, {(\eps \times
10^3)^2 \over M_4^2} < (3 \times 10^{22} \; \hbox{GeV})^{-2} \eeq
which, using $M_4 = 2 \times 10^{18}$ GeV, implies $\eps < 7
\times 10^{-4}$ --- a result which is competitive with the
terrestrial bound, $\eps < 10^{-3}$, when $\eps > 0$, but is not
as good as the bound $\eps \lsim 10^{-6}$ obtained from the
conservation of angular momentum of the sun (see the footnote on
page 5).

\subsubsection{$b_p < b_\gamma$}
If $b_p < b_\gamma$, then the best bound comes from the process
$\gamma \to p\overline{p}$, which does not give a bound even as
good as $\eps < 10^{-3}$ for any $d$ given the constraint $M_d >
1$ TeV.

\subsection{$b_e < b_\gamma$}

The bound when $b_e < b_\gamma$ is not strong enough to be
interesting for $d=4$, so we consider only higher dimensions.
Since the order-of-magnitude result for the electron has $|b_e|
\sim |b_\gamma|$ we have
\eq b_\gamma \sim {\eps^2  \over {(4 \pi)^{[d/2]} \, M_d^2}} \;
\left( {m_t \over M_d} \right)^{d-4}\; \,\left({\alpha \over
4\pi}\right),\eeq
and so, if $b_\gamma - b_e > 0$ the bound becomes:
\eq \eps < \left[ {(4 \pi)^{[d/2]+1}\over {\alpha}} \left( {M_d
\over \ m_t} \right)^{(d-4)} \right]^{1/2} \; \left( { M_d \over
10^{12} \; \hbox{GeV} } \right). \eeq
For $M_d > 1$ TeV this is never better than $\eps < 10^{-6}$, but
it represents progress relative to $\eps < 10^{-3}$ if
\eqa {M_d \over \hbox{TeV}} &<& \left[ { \, 10^{6} \ (
{0.175})^{(d-4)/2} \sqrt{\alpha \over (4 \pi)^{[d/2]+1}}} \,
\right]^{2/(d-2)} \\
&=&  \left\{\matrix{ 90 & (d=5)\cr 10 &(d=6)} \right. \, ,\eeqa
which is only of interest for 5D warped scenarios.

\subsection{Direct Bounds on $b_\gamma$ }
The observation of cosmologically distant gamma rays from
gamma-ray bursters can provide a bound on $|b_\gamma|$ which is
similar in strength to the one just obtained, but applies equally
well to both signs of the result. Ref.~\cite{Ellis} argues that
the absence of dispersion in these signals provides a sensitivity
to changes in the photon dispersion which can be as small as
\eq |b_\gamma| < {( 9\times 10^{11} \; \hbox{GeV})^{-2}},\eeq
which would correspond to
\eq \eps < \left[ {(4 \pi)^{[d/2]+1}\over{\alpha}} \, \left( {M_d
\over \ m_t} \right)^{(d-4)} \right]^{1/2} \; \left( { M_d \over
10^{11} \; \hbox{GeV} } \right). \eeq

This is better than $\eps < 10^{-3}$ provided
 \eqa {M_d \over \hbox{TeV}} &<& \left[ { 10^{5} \,  \ (
{0.175})^{(d-4)/2} \sqrt{{\alpha}\over (4 \pi)^{[d/2]+1}}}
\right]^{2/(d-2)} \nn\\
&=& \left\{ \matrix{ 20 &(d=5) \cr 3 &(d=6)} \right. \eeqa
which is comparable in strength to what was found above.

\section{Conclusions}
Brane-world scenarios allow the possibility of potentially strong
preferred-frame effects arising from extra-dimensional bulk
physics, and suggest these effects may be limited to the graviton
sector. Depending on their sign, the magnitude of such changes to
graviton dispersion can be comparatively large because of the
graviton's extremely weak interactions. Motivated by this
observation, we have explored how gravitational Lorentz-violation
in a brane-world picture influences the properties of observable
particles. We obtain the following results:

\bigskip

{\it 1.} We analyze how Lorentz violation in extra dimensions
arises within the low-energy four-dimensional field theory which
is obtained once the extra dimensions are integrated out. We find
the leading contributions to be the appearance of potentially
different metrics in the kinetic terms which govern the
propagation of the various low-energy particles. If
preferred-frame effects dominate violations of translation or
rotation invariance (in the preferred frame) then in flat space
the metric seen by particle type `$k$' may be written
\eq G_{\mu\nu} = \eta_{\mu\nu} + (1 - c_k^2) \, u_\mu u_\nu \eeq
where $u^\mu$ is the (approximately constant) 4-velocity which
defines the preferred frame. $c_k$ may be interpreted as the
maximum propagation speed of this particle type.

{\it 2.} If subleading contributions at low energies are also
included, then more complicated changes to the dispersion relation
become possible. For low momenta these dispersion relations become
$E_k^2 = p^2 c_k^2 + b_k p^4 + \dots$ in the preferred frame. The
next-to-leading coefficient $b_k$ causes dispersive propagation if
it is nonzero. (Odd powers of $p$, like $p^3$, typically do not
arise in the dispersion relation since they are usually forbidden
by selection rules like rotation invariance in the preferred
frame.)

{\it 3.} Strong observational constraints exist which preclude
large differences between the values $c_k$ and $b_k$ for photons
and other particles. The strongest of these come from the absence
of a dependence on the Earth's velocity through space in atomic
spectra, and from the absence of \v Cerenkov-like decays of
very-high-energy cosmic rays.

{\it 4.} We compute how graviton loops can bring the news of
Lorentz violation in the graviton sector to other particles for
which stronger constraints exist. We do so quite generally at low
energies for the purely massless 4D graviton (the lowest KK mode).
We also compute these loops for the entire KK tower of gravitons,
in the approximation that the dominant contribution comes from
gravitons whose wavelengths are much shorter than are the typical
curvature scales of the extra-dimensional metric.

{\it 5.} We find that one-loop contributions for photons from
graviton-induced Lorentz violation are small. Keeping the finite
and log-divergent parts, we find gravitons do not induce any
change at all in the photon maximum propagation speed, $c_\gamma$.
The purely 4D graviton induces a dispersive term, $b_\gamma \sim
(c_g - 1)^2 /M_p^2$. The contribution of the rest of the KK
graviton tower to the photon energy tends to be suppressed by
further powers of the photon momenta, and so is not important at
low energies. Unfortunately this makes the strong constraints on
photon properties based on gamma-ray bursts and dispersion in
quasar signals \cite{Ellis} largely irrelevant to this kind of
Lorentz violation.

{\it 6.} Using the same approximations, we find fermions acquire
changes to their low-energy dispersion relations with an amount
which varies strongly with the fermion's mass. Among the three
most abundant particles in everyday life -- electrons, protons and
photons -- this predicts by far the biggest effects for protons.
Contributions to $c_k -1 $ are linear in $c_g -1 $ (if $c_g$ is
the graviton maximum speed) while those to $b_k$ are quadratic in
$c_g - 1$.

{\it 7.} The strong mass dependence makes the results very
sensitive to the high-energy spectrum of the theory, since heavy
particles embedded in higher loops can produce larger
contributions to low-energy Lorentz-violating effects than do the
direct graviton loops. An estimate of this effect using the top
quark or $W$-boson as the heavy particle suggests that protons
receive larger contributions than do photons or electrons.

{\it 8.} We find that current atomic constraints on $c_k - 1$ for
observable particles can provide limits on $c_g - 1$ which are
competitive with the direct bound $c_g - 1 \lsim 10^{-6}$ arising
from post-Newtonian corrections in the solar system, but only if
$d=5$ or 6 and if the higher-dimensional Planck mass, $M_d$, is as
close as possible to the TeV scale. For instance, for a warped
$d=5$ model with $M_5 \sim 10$ TeV, we find $|c_g - 1| < 3 \times
10^{-15}\, (M_5/\hbox{TeV})^3$. For $c_g > 1$ this is an
improvement over the direct bound $c_g - 1 < 10^{-6}$ provided
$M_5 < 700$ TeV.

{\it 8.} We find that stronger constraints on $c_g - 1$ arise from
limits on $b_k - b_\gamma$, depending on the dimension of the
extra-dimensional spacetime, and on how low the $d$-dimensional
Planck scale, $M_d$, is. Lower values of $M_d$ lead to better
bounds, which can go up to values of order $M_d \sim 10^7$ TeV
(for $d=5$), provided that $b_p > b_\gamma$. The sign of this
quantity is important, because the bounds which are most
constraining are those which are based on the absence of
too-efficient energy-loss mechanisms for the highest energy cosmic
rays (which we take to be protons), and the decay channel $p \to
p\gamma$ is only open if $b_p > b_\gamma$. Strikingly, purely
4-dimensional graviton loops can give contributions with the right
sign, and which are large enough to produce bounds which are of
order $\eps < 7 \times 10^{-4}$. This makes them comparable with
those from obtained from terrestrial experiments and the binary
pulsar.

\bigskip

Perhaps our most surprising result is that, in some regimes,
graviton loops are already being constrained by observational
data. This is yet another striking way in which the brane-world
picture can run against pre-brane-world intuition.

\vspace{1 cm}
\begin{center}
{\bf Acknowledgements}
\end{center}

We would like to acknowledge fruitful conversations with Debajyoti Chauduri, 
Maxim Pospelov and Xerxes Tata. C.B. wishes to acknowledge
discussions with Detlef Nolte and Witold Skiba at an early point
in this research. J.M. acknowledges discussions with A.Pomarol.
The research of C.B., J.C. and E.F. has been
supported in part by N.S.E.R.C. of Canada and F.C.A.R. of
Qu\'ebec. J.M acknowledges  support by CICYT Research Project
AEN99-0766, MCyT  and the Theory Group of McGill
University.

\section{Appendix A}
In this appendix we record the expressions for the Feynman rule
for the photon-graviton vertex, as obtained from the
Einstein-Maxwell action, eq.\ \pref{DEHAction}. One finds:
\eqa \label{vertices} V^{\alpha\beta:\mu\nu}(P,Q) &=& (P^\mu Q^\nu
+ P^\nu Q^\mu) \, \eta^{\alpha\beta} \nn\\
&& \quad + (P^\beta Q^\alpha - P\cdot Q \, \eta^{\alpha \beta}) \,
\eta^{\mu\nu} \nn\\
&& \quad - (P^\mu Q^\alpha - P\cdot Q \, \eta^{\mu\alpha} ) \,
\eta^{\nu \beta} \nn\\
&& \quad - ( P^\beta Q^\mu - P\cdot Q \, \eta^{\mu\beta} ) \,
\eta^{\nu\alpha} \nn\\
&& \quad - P^\nu Q^\alpha \, \eta^{\mu\beta} - P^\beta Q^\nu \,
\eta^{\mu\alpha} , \eeqa
which has the required symmetry properties:
\eq V^{\alpha\beta:\mu\nu}(P,Q) = V^{\alpha\beta:\nu\mu}(P,Q) =
V^{\beta\alpha:\mu\nu}(Q,P). \eeq

\section{Appendix B}
We here record in more detail the results of the fermion
self-energy calculation, using a graviton loop in $d$ dimensions.

\subsection{d=5}
For five spacetime dimensions evaluation of the graviton loop
gives
\eqa \scA_5 &=& - m^4 \lambda_5 \; \left(\frac{25}{9} + \frac{74}{9}
\, \eps + {85\over 9} \, \eps^2 + ...\right) ,\nn\\
 \scB_5 &=&  - m^3 \lambda_5 \; \left(\frac{35}{18} \, \eps + \frac{839}{180} \, \eps^2 + ...\right) ,\nn\\
 \scC_5 &=& - m^2 \lambda_5  \; \left(\frac{25}{6}\, \eps + \frac{517}{60}\, \eps^2 + ...\right) ,\nn\\
 \scD_5 &=& - m \lambda_5 \; \left(\frac{101}{30} \, \eps^2 + ...\right) ,\nn\\
 \scE_5 &=& -\lambda_5 \; \left(\frac{29}{30} \, \eps^2 + ... \right)\; . \eeqa
where $\lambda_5 = (\kappa_5/8\pi)^2$. The result is finite in
dimensional regularization because one-loop results in odd
dimensions always are with this regularization scheme. This result
agrees with the finite part as computed by directly cutting off
the momentum integrals and ignoring the divergent terms (none of
which are logarithmically divergent).

These lead to the dispersion relation of eq.\ \pref{disprel}:
\eqa c_f^2 -1 &=&  m_f^3 \lambda_5  \; \left[ \frac{110}{9} \,
\eps + \frac{239}{9} \, \eps^2 + ... \right], \nn\\
b_f &=&   m_f \, \lambda_5 \left[ \frac{26}{3} \, \eps^2 + ...
\right] \eeqa
Notice that both of these results are positive (provided $\eps >
0$).

\subsection{d=6}
For six spacetime dimensions the graviton loop gives
\eqa \scA_6 &=& - m^5 \lambda_6 \; \left(\frac{9}{4} + \frac{25}{4}
\, \eps + {53\over 8} \, \eps^2 + ...\right) ,\nn\\
 \scB_6 &=& - m^4 \lambda_6 \; \left(\frac{6}{5} \, \eps + \frac{83}{30} \, \eps^2 + ...\right) ,\nn\\
 \scC_6 &=& - m^3 \lambda_6  \; \left(\frac{18}{5}\, \eps + \frac{107}{15}\, \eps^2 + ...\right) ,\nn\\
 \scD_6 &=& - m^2 \lambda_6 \; \left(\frac{34}{15} \, \eps^2 + ...\right) ,\nn\\
 \scE_6 &=& - m \lambda_6 \; \left(\frac{4}{3} \, \eps^2 + ... \right)\; . \eeqa
where $\lambda_6 = 2\kappa_6^2/(8\pi)^3 \, {\cal L}$. As before
${\cal L} = \log(\Lambda^2/\mu^2) = 2/(6-n)$, when evaluated with
an ultraviolet cutoff, $\Lambda$, and in dimensional
regularization.

These lead to the $d=6$ results:
\eqa c_f^2 -1 &=&  m_f^4 \, \lambda_6  \; \left[ \frac{48}{5}
\, \eps + \frac{99}{5} \, \eps^2 + ... \right], \nn\\
b_f &=&  m_f^2 \, \lambda_6 \left[ \frac{36}{5} \, \eps^2 + ...
\right] \eeqa
Again both results are positive for $\eps > 0$.

\subsection{d=7}
Next, seven spacetime dimensions:
\eqa \scA_7 &=&  m^6 \lambda_7 \; \left(\frac{49}{25} +
\frac{26}{5} \, \eps + {129\over 25} \, \eps^2 + ...\right) ,\nn\\
 \scB_7 &=&  m^5 \lambda_7 \; \left(\frac{21}{25} \, \eps + \frac{131}{70} \, \eps^2 + ...\right) ,\nn\\
 \scC_7 &=&  m^4 \lambda_7  \; \left(\frac{49}{15}\, \eps + \frac{263}{42}\, \eps^2 + ...\right) ,\nn\\
 \scD_7 &=&  m^3 \lambda_7 \; \left(\frac{59}{35} \, \eps^2 + ...\right) ,\nn\\
 \scE_7 &=&  m^2 \lambda_7 \; \left(\frac{53}{35} \, \eps^2 + ... \right)\; . \eeqa
where $\lambda_7 = \kappa_7^2/[6(4\pi)^3]$, and is finite when
evaluated in dimensional regularization.

These lead to the $d=7$ results:
\eqa c_f^2 -1 &=& - m_f^5 \, \lambda_7  \; \left[ \frac{616}{75}
\, \eps + \frac{244}{15} \, \eps^2 + ... \right], \nn\\
b_f &=& - m_f^3 \, \lambda_7 \left[ \frac{32}{5} \, \eps^2 + ...
\right] \eeqa
Here both results are negative for $\eps > 0$.

\subsection{d=8}
For eight spacetime dimensions we have
\eqa \scA_8 &=&  m^7 \lambda_8 \; \left(\frac{16}{9} +
\frac{41}{9}
\, \eps + {77\over 18} \, \eps^2 + ...\right) ,\nn\\
 \scB_8 &=&  m^6 \lambda_8 \; \left(\frac{40}{63} \, \eps + \frac{173}{126} \, \eps^2 + ...\right) ,\nn\\
 \scC_8 &=&  m^5 \lambda_8  \; \left(\frac{64}{21}\, \eps + \frac{239}{42}\, \eps^2 + ...\right) ,\nn\\
 \scD_8 &=&  m^4 \lambda_8 \; \left(\frac{4}{3} \, \eps^2 + ...\right) ,\nn\\
 \scE_8 &=&   m^3 \lambda_8 \; \left(\frac{34}{21} \, \eps^2 + ... \right)\; . \eeqa
where $\lambda_8 = 2\kappa_8^2/(8\pi)^4 \, {\cal L}$. As before
${\cal L} = \log(\Lambda^2/\mu^2) = 2/(8-n)$, when evaluated with
an ultraviolet cutoff, $\Lambda$, and in dimensional
regularization.

These lead to the $d=8$ results:
\eqa c_f^2 -1 &=& - m_f^6 \, \lambda_8  \; \left[ \frac{464}{63}
\, \eps + \frac{890}{63} \, \eps^2 + ... \right], \nn\\
b_f &=& -  m_f^4 \, \lambda_8 \left[ \frac{124}{21} \, \eps^2 + ...
\right] \eeqa
Here both results are negative for $\eps > 0$.

\subsection{d=9}
For $d=9$ we have
\eqa \scA_9 &=&  -\,m^8 \lambda_9 \; \left(\frac{81}{49} +
\frac{202}{49}
\, \eps + {181\over 49} \, \eps^2 + ...\right) ,\nn\\
 \scB_9 &=& -\, m^7 \lambda_9 \; \left(\frac{99}{196} \, \eps +
 \frac{3755}{3528} \, \eps^2 + ...\right) ,\nn\\
 \scC_9 &=& -\, m^6 \lambda_9  \; \left(\frac{81}{28}\, \eps +
 \frac{2665}{504}\, \eps^2 + ...\right) ,\nn\\
 \scD_9 &=& -\, m^5 \lambda_9 \; \left(\frac{277}{252} \, \eps^2 + ...\right) ,\nn\\
 \scE_9 &=& -\, m^4 \lambda_9 \; \left(\frac{425}{252} \, \eps^2 + ... \right)\; . \eeqa
where $\lambda_9 = \kappa_9^2/[15 (4\pi)^4]$.

These lead to the results:
\eqa c_f^2 -1 &=&  m_f^7 \, \lambda_9  \; \left[ \frac{333}{49}
\, \eps + \frac{1245}{98} \, \eps^2 + ... \right], \nn\\
b_f &=&  m_f^5 \, \lambda_9 \left[ \frac{39}{7} \, \eps^2 + ...
\right] \eeqa
Here both results are positive for $\eps > 0$.

\subsection{d=10}
Finally, for ten spacetime dimensions:
\eqa \scA_{10} &=& -\,m^9 \lambda_{10} \; \left(\frac{25}{16} +
\frac{61}{16}
\, \eps + {105\over 32} \, \eps^2 + ...\right) ,\nn\\
 \scB_{10} &=& -\,m^8 \lambda_{10} \; \left(\frac{5}{12} \, \eps +
 \frac{103}{120} \, \eps^2 + ...\right) ,\nn\\
 \scC_{10} &=& -\, m^7 \lambda_{10}  \; \left(\frac{25}{9}\, \eps +
 \frac{449}{90}\, \eps^2 + ...\right) ,\nn\\
 \scD_{10} &=& -\, m^6 \lambda_{10} \; \left(\frac{14}{15} \, \eps^2
 + ...\right) ,\nn\\
 \scE_{10} &=& -\, m^5 \lambda_{10} \; \left(\frac{26}{15} \, \eps^2
 + ... \right)\; . \eeqa
where $\lambda_{10} = 4\kappa_{10}^2/[3(8\pi)^5] \, {\cal L}$, and
${\cal L} = \log(\Lambda^2/\mu^2) = 2/(10-n)$, when
evaluated with an ultraviolet cutoff, $\Lambda$, and in
dimensional regularization.

These lead to the $d=10$ results:
\eqa c_f^2 -1 &=&  m_f^8 \, \lambda_{10} \; \left[
\frac{115}{18} \, \eps + \frac{421}{36} \, \eps^2 + ... \right], \nn\\
b_f &=&  m_f^6 \, \lambda_{10} \left[ \frac{16}{3} \, \eps^2 +
... \right] \eeqa
$\eps > 0$ ensures that both results in this case are positive.

\end{document}